\documentclass[english,a4paper]{article}
\usepackage{fullpage}
\usepackage{amssymb}
\usepackage{amsmath}
\usepackage{url}
\usepackage{ulem}
\usepackage{bm}
\usepackage{color}
\usepackage{booktabs}
\usepackage{graphicx}
\usepackage{authblk}
\usepackage{pdfpages}
\usepackage{graphicx}
\usepackage{amsfonts}
\usepackage{hyperref}

\DeclareMathOperator{\Tr}{Tr}
\DeclareMathOperator{\var}{var}

\DeclareMathOperator{\E}{E}
\DeclareMathOperator{\bias}{bias}
\DeclareMathOperator{\diag}{diag}

\begin{document}
\date{}
\title{DRAGON: Determining Regulatory Associations using Graphical models on multi-Omic Networks}

\author[1,2,$\ddagger$]{Katherine H. Shutta}
\author[1,$\dagger\ddagger$]{Deborah Weighill}
\author[6]{Rebekka Burkholz}
\author[1]{Marouen Ben Guebila}
\author[2]{Dawn L. DeMeo}
\author[3,4]{Helena U. Zacharias}
\author[1,2,$*$]{John Quackenbush}
\author[1,5,$*$]{Michael Altenbuchinger}

\affil[1]{Department of Biostatistics, Harvard T.H. Chan School of Public Health, Boston, MA, USA}
\affil[2]{Channing Division of Network Medicine, Department of Medicine, Brigham and Women's Hospital and Harvard Medical School, Boston, MA, USA}
\affil[3]{Department of Internal Medicine I, University Medical Center Schleswig-Holstein, Campus Kiel, Kiel, Germany}
\affil[4]{Institute of Clinical Molecular Biology, Kiel University and University Medical Center Schleswig-Holstein, Campus Kiel, Kiel, Germany}
\affil[5]{Department of Medical Bioinformatics, University Medical Center G\"ottingen, G\"ottingen, Germany}
\affil[6]{CISPA Helmholtz Center for Information Security, Saarbr\"ucken, Germany}
\affil[$\dagger$]{Current address: Lineberger Comprehensive Cancer Center, University of North Carolina at Chapel Hill, Chapel Hill, NC, USA}
\affil[$\ddagger$]{authors contributed equally to this work}
\affil[$*$]{shared senior authors}

\maketitle
\begin{abstract}{The increasing quantity of multi-omics data, such as methylomic and transcriptomic profiles, collected on the same specimen, or even on the same cell, provide a unique opportunity to explore the complex interactions that define cell phenotype and govern cellular responses to perturbations. We propose a network approach based on Gaussian Graphical Models (GGMs) that facilitates the joint analysis of paired omics data. This method, called DRAGON (Determining Regulatory Associations using Graphical models on multi-Omic Networks), calibrates its parameters to achieve an optimal trade-off between the network's complexity and estimation accuracy, while explicitly accounting for the characteristics of each of the assessed omics ``layers.'' In simulation studies, we show that DRAGON adapts to edge density and feature size differences between omics layers, improving model inference and edge recovery compared to state-of-the-art methods. We further demonstrate in an analysis of joint transcriptome - methylome data from TCGA breast cancer specimens that DRAGON can identify key molecular mechanisms such as gene regulation via promoter methylation. In particular, we identify Transcription Factor AP-2 Beta (TFAP2B) as a potential multi-omic biomarker for basal-type breast cancer. DRAGON is available as open-source code in Python through the Network Zoo package (netZooPy v0.8; netzoo.github.io).}
\end{abstract}
\newpage
\section{Introduction}
\enlargethispage{-65.1pt}
Many biological systems can be visualized using networks, where biologically relevant elements are represented as nodes and relationships between those elements are represented as edges. Examples include gene regulatory networks, which represent the regulation of genes by transcription factors, and protein--protein interaction networks, which capture physical interactions between proteins \cite{stelzl2005human,rual2005towards}. 

Network models can be based on prior knowledge \cite{szklarczyk2019string}, inferred from data \cite{markowetz2007inferring}, or combinations thereof \cite{glass2013passing}. Here, we focus on data-driven network inference from high-throughput multi-omic data. In this context, co-expression networks \cite{aoki2007approaches}, which are based on a measure of correlation such as Pearson's correlation, are often used to capture potential associations between biomolecules that may be coordinately altered in specific biological states. However, a major drawback of such networks is that they do not distinguish direct from indirect effects \cite{altenbuchinger2020gaussian}. For example, consider a situation where a transcription factor $A$ regulates the expression of two genes, $B$ and $C$. In this case, a correlation network will contain an edge between gene $B$ and gene $C$ because correlation indicates a relation between the two genes. However, that relationship is only a consequence of their mutual relationship with the transcription factor $A$ and thus the observed edge in the correlation network represents an indirect association. The problem of such erroneous correlations was discussed by Pearson and Yule in the early 20th century, where the term ``spurious correlation'' was introduced to distinguish indirect from direct relationships. A historical review of this question has been summarized by Aldrich and colleagues \cite{aldrich1995correlations}.

Several approaches have attempted to address this issue \cite{margolin2006aracne,butte1999mutual,wille2004sparse,schafer2005shrinkage}, of which Gaussian graphical models (GGMs; also known as partial correlation networks) \cite{wille2004sparse,schafer2005shrinkage} are among the most widely used methods. In a GGM, edges represent partial correlations. Intuitively, the partial correlation between two variables can be considered as the correlation that takes into account the effect of all remaining variables in the data set. Thus, it can distinguish a direct relationship from one that is mediated by one or more other variables.  GGMs outperform simple correlation networks \cite{krumsiek2011gaussian} and were consistently among the best in comparison to other methods in finding meaningful associations \cite{ghanbari2019distance}. 

A single type of omics data generally only provides part of the information necessary to distinguish between direct and incidental relationships. For example, we know that gene regulation is a process that involves multiple layers of control, including transcription factor binding, epigenetic regulation, and chromatin structure. However, many analyses incorporate only gene expression data and not other regulatory data. Incorporating data from multiple omics could help prevent possibly erroneous conclusions based on the concept of spurious correlations introduced above. 

New technologies are making it possible to generate multiple layers of omics data from the same samples. For example, The Cancer Genome Atlas (TCGA) provides data on RNA expression, methylation levels, and copy number variations for many individual tumor samples. More recently, single cell multi-omic data have become available as it has become possible to assay different omics data types from individual cells; for example, Cao et al. (2018) measured RNA and chromatin accessibility in single cells\cite{cao2018joint}. Such multifactorial data will allow us to better disentangle interactions between biological variables and distinguish genuine from spurious associations. 

Most omics data are high dimensional, meaning that the number of measured variables $p$ typically exceeds the number of samples $n$ (or both are of the same order of magnitude), which presents challenges for network inference \cite{altenbuchinger2020gaussian}. Several remedies have been proposed based on regularization techniques from high-dimensional statistics \cite{wille2004sparse,schafer2005shrinkage, meinshausen2006high,friedman2008sparse}. Multi-omics network inference is complicated by a number of factors including the larger numbers of variables, different numbers of variables for each omics layer, variable noise levels within and between layers, and different edge densities in each data type. 

In this work, we propose \textbf{DRAGON} (\textbf{D}etermining \textbf{R}egulatory \textbf{A}ssociations using \textbf{G}raphical models on multi-\textbf{O}mic \textbf{N}etworks), a machine learning method to estimate GGMs using two omics layers simultaneously. DRAGON calibrates omic-specific parameters for each omic layer to achieve an optimal trade-off between model complexity and estimation accuracy while explicitly taking into account the unique characteristics of each omics layer.

We show in simulations that DRAGON adapts to differences in edge densities and feature sizes of the included omics layers. Finally, we use DRAGON to analyze joint transcriptome--methylome data from TCGA breast cancer specimens. The latter analysis shows that DRAGON can identify potential regulatory molecular mechanisms, such as the association between promoter methylation and gene expression. {We further show that DRAGON can identify multi-omic biomarkers, as exemplified by the combination of promoter methylation and gene-expression of TFAP2B (Transcription Factor AP-2 Beta), which is strongly associated with the basal-like breast cancer subtype. }

\section{Materials and Methods}

\subsection{The Gaussian Graphical Model}
Let ${X}$ be a $n\times p$ data matrix of $n$ observations (samples) and $p$ features (such as genes, methylated sites, or proteins). Assume that the observations $\mathbf{x}_1,\mathbf{x}_2,\ldots, \mathbf{x}_n \in \mathbb{R}^p$ are independent and identically distributed according to a multivariate normal, $N({\mu},{\Sigma})$, where ${\Sigma}$ is a positive definite covariance matrix. Further, ${\Theta}=(\theta_{ij})={\Sigma}^{-1}$ is the inverse covariance matrix (also called precision matrix) where vanishing entries $\theta_{ij}$ correspond to conditional independencies between variables $i$ and $j$. A Gaussian Graphical Model (GGM) is a conditional dependence graph in which nodes represent variables and edges connect conditionally dependent pairs of variables \cite{lauritzen1996graphical,bishop2006pattern}.

Let ${S}=\frac{1}{n}\sum_{i=1}^n(\mathbf{x}_i-\hat{\mu})(\mathbf{x}_i-\hat{\mu})^T$ be the sample covariance matrix, where $\hat \mu=\frac{1}{n}\sum_{i=1}^n \mathbf{x}_i$ is the sample mean. Then, the corresponding log likelihood takes the form
\begin{equation}
l({\Theta})= \frac{n}{2}\left(\log |{\Theta}|-\Tr\left({S}{\Theta}\right)\right).
 \label{MLE_Gauss}
\end{equation}

\subsection{Covariance shrinkage}
The Maximum Likelihood Estimate (MLE) of Eq. (\ref{MLE_Gauss}) yields $\hat{\Theta} = {S}^{-1}$. However, if the number of features $p$ exceeds the number of independent observations $n$, then ${S}$ is singular and cannot be inverted. Even if $p$ is smaller than $n$ but of the same order of magnitude, $\hat{\Theta}$ has a high variance. One way this issue is often addressed by adding regularization terms to Eq. (\ref{MLE_Gauss}), as for example proposed in \cite{friedman2008sparse}. Another approach is that of Sch\"afer and colleagues \cite{schafer2005shrinkage}, who bias the covariance matrix towards a target matrix that is typically full-rank. Such ``covariance shrinkage" is based on the biased estimator
\begin{equation}
{\hat{\Theta}} = \left((1-\lambda){S}+\lambda {T}\right)^{-1}\,,
\label{CovShrink}
\end{equation}
where $\lambda\in[0,1]$ is a regularization parameter that can be estimated using the Lemma of Ledoit \& Wolf \cite{ledoit2000well} and ${T}$ is the target matrix. Here, different choices for ${T}$ have been proposed, such as the identity, ${T}={I}_{p}$, and the diagonal of ${S}$, ${T}=\diag (s_{11},s_{22},\ldots,s_{pp})$. {The idea behind Eq. (\ref{CovShrink}) is to replace the empirical covariance $S$ in the MLE of $\Theta$ by a linear combination of $S$ and the target matrix $T$ representing conditional independence. Because $T$ is full-rank, the singularity of $S$ is mitigated in this sum. Consequently, a biased precision matrix estimator can be obtained from inverting the shrunken covariance matrix.} Throughout this article, we use the target matrix {${T}=\diag (s_{11},s_{22},\ldots,s_{pp})$} following the arguments in \cite{schafer2005shrinkage}. 

\subsection{DRAGON}
\subsubsection{Generalized covariance shrinkage}
In DRAGON, we extend covariance shrinkage to account for two different omics layers by introducing layer-specific regularization terms. Let ${X}^{(1)}$ be a $n\times p_1$ data matrix that represents the first omics layer and let ${X}^{(2)}$ be a $n\times p_2$ data matrix for the second layer, where $p_1$ and $p_2$ are the number of variables from omics layers $1$ and $2$. We further assume paired data, meaning that measurements (rows) $\mathbf{x}_i^{(1)}$ and $\mathbf{x}_i^{(2)}$ correspond to the same sample $i$ but differ in their features. We define the empirical covariances $S^{(k, l)}=\frac{1}{n}\sum_{i=1}^n(\mathbf{x}_i^{(k)}-{\hat{\mu}}^{(k)})(\mathbf{x}_i^{(l)}-{\hat{\mu}}^{(l)})^T$ with the empirical mean vector $\hat{\mu}^{(k)}=\frac{1}{n}\sum_{i=1}^n \mathbf{x}_i^{(k)}$ for $k,l\in \{1,2\}$. Now, we can generalize the shrinkage estimator to
\begin{eqnarray}
{\hat{\Theta}}&=&\Bigg(\left(\begin{array}{cccc}(1-\lambda_1) {S}^{(1,1)}&\sqrt{1-\lambda_1}\sqrt{1-\lambda_2} S^{(1,2)}\\\sqrt{1-\lambda_1}\sqrt{1-\lambda_2} S^{(2,1)}&(1-\lambda_2) S^{(2,2)} \end{array}\right)\nonumber\\&&+\left(\begin{array}{cc}\lambda_1  T^{(1)}&0\\0&\lambda_2  T^{(2)}\end{array}\right)\Bigg)^{-1}\,,
\label{DRAGON}
\end{eqnarray}
with $\lambda_k\in[0, 1]$ and ${T}^{(k)} = \textrm{diag}(s^{(k)}_{11}, s^{(k)}_{22}, \ldots, s^{(k)}_{p_k p_k})$, where ${S}^{(k,k)}=(s^{(k)}_{ij})$. For illustration purposes, we first consider the limit $\lambda_2=1$. Here, Eq. (\ref{DRAGON}) becomes
\begin{eqnarray}
{\hat{\Theta}}&=&\left(\left(\begin{array}{cccc}(1-\lambda_1)  S^{(1,1)}&0\\0&0 \end{array}\right)+\left(\begin{array}{cc}\lambda_1  T^{(1)}&0\\0&  T^{(2)}\end{array}\right)\right)^{-1}\nonumber\\
&=&\left(\begin{array}{cccc} \Theta^{(1,1)}&0\\0&\left(  T^{(2)}\right)^{-1} \end{array}\right)\,,
\end{eqnarray}
where $ \Theta^{(1,1)}=\left((1-\lambda_1)  S^{(1,1)} + \lambda_1  T^{(1)}\right)^{-1}$ is the shrinkage estimator of the precision matrix using only the features with $k=1$. Thus, if $\lambda_2=1$, technology $1$ decouples from technology $2$. 

Next, consider the limit $\lambda_1=\lambda_2=\lambda$. Then Eq. (\ref{DRAGON}) becomes
\begin{eqnarray}
{\hat{\Theta}}&=&\left((1-\lambda)\left(\begin{array}{cccc} S^{(1,1)}& S^{(1,2)}\\ S^{(2,1)}& S^{(2,2)} \end{array}\right)+\lambda\left(\begin{array}{cc}  T^{(1)}&0\\0& T^{(2)}\end{array}\right)\right)^{-1}\nonumber\\
&=&\left((1-\lambda) S+\lambda  T\right)^{-1}\,.
\end{eqnarray}
Thus we naively treat both omics layers as if they were generated using the same technology. These examples show that DRAGON naturally incorporates two limits that bound the optimal solution: (i) GGMs estimated for the two omics layers separately and (ii) a GGM treating both layers such as if they belong to the same layer.

\subsubsection{Generalization of the Lemma of Ledoit \& Wolf}
The penalty parameters $\lambda_1$ and $\lambda_2$ can be estimated using cross-validation or resampling; however, such approaches are computationally expensive. Alternatively, one can use an analytical estimate following the arguments of Ledoit \& Wolf \cite{ledoit2000well}. There, the shrinkage parameter $\lambda$ was derived by minimizing
\begin{equation}
R=\E\!\left[|| {\hat{\Sigma}}-{\Sigma}||_F^2\right]\,
\label{LedWol}
\end{equation}
with respect to $\lambda$, where ${\hat{\Sigma}}= (1-\lambda){S} + \lambda {T}$ and ${\Sigma}$ is the true, underlying covariance. This is possible since $\bias ({S}) = 0$ makes Eq. (\ref{LedWol}) independent of ${\Sigma}$.

Here, we extend this approach to the shrinkage formula (\ref{DRAGON}) and estimate $\lambda_1$ and $\lambda_2$ by minimizing
\begin{equation}
\E\!\left[||  {\hat{\Sigma}}-{\Sigma}||_F^2\right]=\sum_{k,l=1}^2\E\!\left[|| {\hat{\Sigma}}^{(k,l)}-{{\Sigma}}^{(k,l)}||_F^2\right]
\end{equation}
with respect to $\lambda_1$ and $\lambda_2$, where $||.||_F$ is the Frobenius norm. Following the arguments in \cite{schafer2005shrinkage} {(see Suppl. Section 1)}, we obtain 
\begin{eqnarray}
R&=&\E\!\left[|| {\hat{\Sigma}}-{\Sigma}||_F^2\right] \nonumber\\
&=&\textrm{const.} + \lambda_1 T^{(1)}_{1}+ \lambda_2 T^{(2)}_{1}+ \lambda_1^2 T^{(1)}_{2}+ \lambda_2^2 T^{(2)}_{2} \nonumber\\
&&+ \lambda_1\lambda_2 T_{3}+ \sqrt{1-\lambda_1}\sqrt{1-\lambda_2}T_{4}\label{Rform}
\end{eqnarray}
where the constant term is independent of $\lambda_1$ and $\lambda_2$, and
\begin{eqnarray}
 T^{(k)}_{1}&=&-2\left(\sum_{i\ne j}\var(s^{(k)}_{ij})+\sum_{i, j}\E((s^{(1,2)}_{ij})^2)\right)\,,\nonumber\\
 T^{(k)}_{2}&=&\sum_{i\ne j} \E((s^{(k)}_{ij})^2)\,,\nonumber\\
 T_3&=&2\sum_{i, j}\E((s^{(1,2)}_{ij})^2)\,,\nonumber\\
 T_4&=&4\sum_{i, j}\left(\var(s^{(1,2)}_{ij})-\E((s^{(1,2)}_{ij})^2)\right)\,\nonumber
\end{eqnarray}
Eq. (\ref{Rform}) can be easily minimized with respect to $\lambda_1\in [0,1]$ and $\lambda_2\in [0,1]$, where the moments can be estimated following \cite{schafer2005shrinkage}. 

\subsubsection{Hypotheses testing}
\label{SigTest}
An estimate for the partial correlation between variable $i$ and $j$ can be directly obtained from ${\hat{\Theta}}=( \hat\theta_{ij})$ by calculating
\begin{equation}
\hat \rho_{ij} = -\frac{\hat \theta_{ij}}{\sqrt{\hat \theta_{ii}\hat\theta_{jj}}}\,.
\label{norm}
\end{equation}
As a consequence of covariance shrinkage, the partial correlation matrix ${\hat{P}}=(\hat \rho_{ij})$ is also shrunken \cite{bernal2019exact}. Bernal et al. (2019) developed a null-model probability density that naturally accounts for this shrinkage effect \cite{bernal2019exact}:

\begin{equation}
f^\lambda_{0}(\rho)=\frac{\left((1-\lambda)^2-\rho^2\right)^{(\kappa-3)/2}}{\textrm{Beta}(\frac{1}{2},\frac{\kappa-1}{2}) (1-\lambda)^{\kappa-2}}\,,
\label{Null}
\end{equation}   
where the parameter $\kappa$ is given by $n-1-(p-2)$ for $n\gg p$, or can be fitted by MLE for the ill-posed case $p<n$ or for $p\approx n$ \cite{bernal2019exact}. {For an intuitive derivation of Eq. (\ref{Null}) see \cite{bernal2019exact}.} Let
\begin{equation}
{\hat{\Theta}}=\left(\begin{array}{cccc}{\hat{\Theta}}^{(1,1)}&{\hat{\Theta}}^{(1,2)}\\{\hat{\Theta}}^{(2,1)}&{\hat{\Theta}}^{(2,2)} \end{array}\right)\quad\mbox{and}\quad {\hat{P}}=\left(\begin{array}{cccc}{\hat{P}}^{(1,1)}&{\hat{P}}^{(1,2)}\\{\hat{P}}^{(2,1)}&{\hat{P}}^{(2,2)} \end{array}\right)\,,
\end{equation}
where ${\hat{\Theta}}^{(1,1)}$ (${\hat{P}}^{(1,1)}$) has dimension $p_1\times p_1$ and ${\hat{\Theta}}^{(2,2)}$ (${\hat{P}}^{(2,2)}$) dimension $p_2\times p_2$. Then, DRAGON assigns significance levels to partial correlations using the following steps:
\begin{itemize}
\item[(i)] Simulate data under the null hypotheses $(H_0: \rho = 0)$ for given sample size $n$, and estimate corresponding partial correlations using DRAGON with $\lambda_1$ and $\lambda_2$ given from the original data. 
\item[(ii)] Fit $\kappa$ using MLE of Eq. (\ref{Null}) for ${{P}}^{(1,1)}$, ${{P}}^{(1,2)}$, and ${{P}}^{(2,2)}$, separately.
\item[(iii)] Use density Eq. (\ref{Null}) with $\kappa$ determined in (ii) to assign significance levels to ${\hat{P}}^{(1,1)}$, ${\hat{P}}^{(1,2)}$, ${\hat{P}}^{(2,2)}$, respectively.
\item[(iv)] Adjust significance levels in layer $(1,1)$, $(1,2)$, and $(2,2)$ for multiple testing, separately, using the method of Benjamini and Hochberg \cite{benjamini1995controlling}. Note that if we control the false discovery rate (FDR) at level $\alpha$ in each layer, we also control the overall FDR at level $\alpha$ across all layers. However, to provide good estimates, $p_1$ and $p_2$ must be assumed to be sufficiently large.
\end{itemize}
For $n\gg p$, $\kappa$ is given by $\kappa=n-1-(p-2)$ and (i -- ii) are not necessary.

\subsection{{Performance comparisons}}
{To benchmark the performance of DRAGON, we selected five methods for GGM estimation, each of which has distinct advantages and disadvantages \cite{altenbuchinger2020gaussian}, based on the requirement that they are available through a user-friendly software, provide estimates for $p$-values without computationally expensive resampling, and have been published in peer-reviewed journals.}
\begin{itemize}
\item{\textbf{GGM:} we implemented an omics-layer agnostic DRAGON model, which we simply denote as Gaussian Graphical Model (GGM) in the following. Note, in contrast to the GeneNet implementation described below \cite{GeneNet}, this approach uses the exact null distribution for shrunken partial correlations as suggested by \cite{bernal2019exact}.}
\item{\textbf{GeneNet:} the R-package ``GeneNet'' \cite{GeneNet} uses covariance shrinkage and provides estimates for adjusted \textit{p}-values (\textit{q}-values) via an empirical Bayes approach \cite{schafer2005empirical}. We used the standard settings for all comparisons.} 
\item{\textbf{B-NW-SL:} the bivariate nodewise scaled lasso was suggested by \cite{ren2015asymptotic} and uses a regression approach to obtain asymptotically efficient estimates of the precision matrix under a sparseness condition relative to the sample size. We used the implementation provided in the ``SILGGM'' R-package \cite{zhang2018silggm} with standard settings.} 
\item{\textbf{D-S-NW-SL:} the de-sparsified nodewise scaled lasso \cite{jankova2017honest} uses a modification of the nodewise lasso regression approach suggested by \cite{meinshausen2006high} with a de-sparsified estimator. We used the ``SILGGM'' R-package \cite{zhang2018silggm} with standard settings. }
\item{\textbf{D-S-GL:} the de-sparsified graphical lasso was proposed by \cite{jankova2015confidence} and is based on a de-sparsified modification of the graphical lasso proposed by \cite{friedman2008sparse}. We used the ``SILGGM'' R-package \cite{zhang2018silggm} with standard settings.}
\end{itemize}
{For all methods except GeneNet, which provides adjusted $p$-values ($q$-values), we adjusted \textit{p}-values using the procedure of Benjamini and Hochberg \cite{benjamini1995controlling}.}

\section{Results and Discussion}
\subsection{Simulation studies}
{We begin by briefly reviewing the concept of partial correlations and their relevance for multi-omics data analysis. Subsequently, we use four different simulation studies to show a comprehensive performance comparison between DRAGON and competing methods for GGM estimation.} 

\begin{figure}[t!]
\centerline{\includegraphics[width=0.9\textwidth]{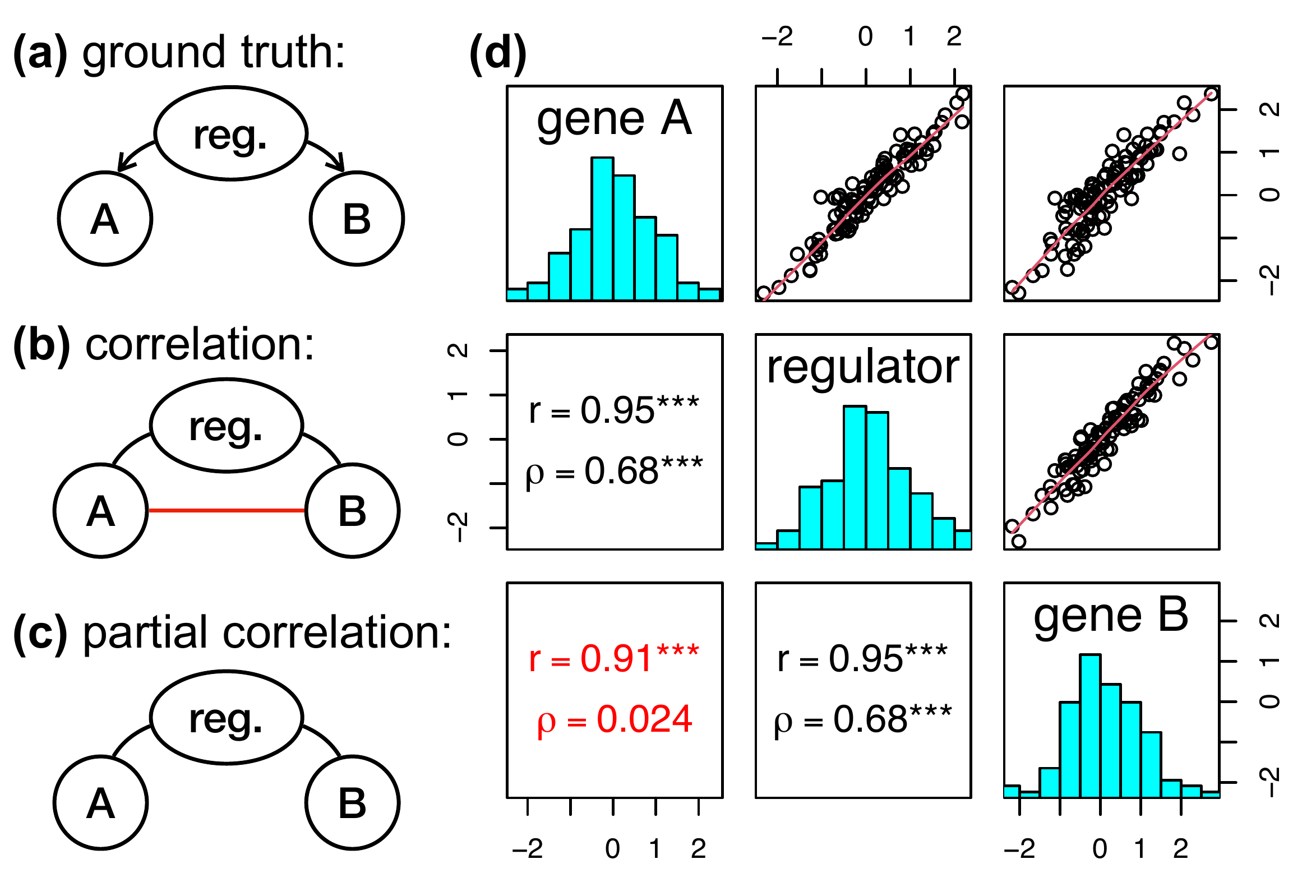}}
\caption{{Partial correlation versus Pearson's correlation on artificial data. The left figure shows from top to bottom: (a) the ground truth (directed), (b) the inferred correlation network (undirected), and (c) the inferred partial correlation network (undirected). Direct relationships are shown in black. Pearson's correlation erroneously infers a relationship between $A$ and $B$ (Fig. a, red edge). Partial correlation correctly removes this relationship (Fig. c). The right figure (Fig. d) shows the corresponding data as scatterplots (upper half of the matrix) and histograms (diagonal). Corresponding Pearson's correlation, $r$, and partial correlations, $\rho$, are given in the lower triangular matrix. Significance levels $p<0.001$ are marked by three asterisks.}} 
\label{Fig_par_cor}
\end{figure}

{Multiple layers of interacting regulatory processes are involved in determining a cell's state. By considering molecular variables of a single omic layer, such as the transcriptome, we can miss relevant information which might lead to erroneous conclusions about causal associations. To illustrate this effect, consider a scenario where the transcription of two genes, $A$ and $B$, is regulated via the same molecular (\textit{regulator}) ``variable''. The \textit{regulator} might belong to another omics layer representing a different biological factor such as the chromatin state in the DNA region of $A$ and $B$. We generated artificial data representing such a process for $n=100$ observations (Suppl. Section 2, Fig.\ \ref{Fig_par_cor}d), assuming a linear relationship between the observed level of the \textit{regulator} and that of gene $A$, and between the \textit{regulator} and gene $B$. Fig.\ \ref{Fig_par_cor}b shows the corresponding correlation network and Fig. \ \ref{Fig_par_cor}d shows the estimated Pearson correlations, $r$, showing a high correlation between $A$ and $B$  (in red, $r\sim 0.91$ and $p<0.001$). This correlation is statistically significant although gene $A$ and gene $B$ are not directly related -- rather, they are only co-regulated by the \textit{regulator}. In contrast, the partial correlation between $A$ and $B$ takes into account the effect of the \textit{regulator}, resulting in a non-significant partial correlation of $\rho\sim0.02$ (Fig. \ref{Fig_par_cor}c, using standard partial correlation) and a partial correlation network that does not have an edge between $A$ and $B$ (Fig. \ref{Fig_par_cor}c), better reflecting the ground truth (Fig. \ref{Fig_par_cor}a). Thus, if the \textit{regulator} is included in the model, it is statistically possible to disentangle direct and indirect linear relationships.}

{
In this section, we systematically analyze potential issues in estimating partial correlations between variables of different omics layers, and compare how different methods, including DRAGON, are affected by these issues. In our simulations, we considered both high-dimensional settings (many molecular variables) as well as models in which there are different probabilities for direct relationships among variables within and between different omics layers. Table \ref{simstud} describes the simulation study design in detail.}

{We performed four simulation studies, $A$ to $D$, with fixed numbers of molecular variables, $p_1=100$ and $p_2=500$ for omics layer 1 and 2, respectively. Direct relationships can be simulated by appropriate choice of the entries of the precision matrices ${\hat{\Theta}}^{(k,l)}$ which parameterize the multivariate Gaussian distribution. Different simulation runs correspond to different precision matrices, which were randomly generated as follows:} 

\begin{itemize}
\item[(0)] start with a $p$-dimensional identity matrix, where $p=p_1+p_2$, 
\item[(1)] randomly replace a proportion $\eta^{(k,l)}$ of zeros by values drawn from a uniform distribution ranging from $-1$ to $1$,
\item[(2)] replace the diagonal entries $\theta_{ii}$ by $\sum_{j}|\theta_{ij}|$ plus a small positive value ($\epsilon=0.0001$) so that the precision matrix is invertible,
\item[(3)] normalize the entries $\theta_{ij}$ of the precision matrix by $\sqrt{\theta_{ii}\theta_{jj}}$.
\end{itemize}
Each precision matrix was used to sample from a multivariate normal with mean vector $\mu=0$ and covariance ${\hat{\Theta}}^{-1}$. Finally, we added a noise $\epsilon\sim N(\mu=0, \sigma=0.1)$ to each entry of the data matrix.
\begin{table}
\centering
\caption{\label{simstud}Edge densities and edge numbers for simulation studies $A$, $B$, $C$, and $D$.}
\begin{tabular}{|l|l|l|l|l|l|l|}
\hline
\multicolumn{1}{|c|}{$(k,l)$} & \multicolumn{2}{c|}{$(1,1)$} & \multicolumn{2}{c|}{$(1,2)$} & \multicolumn{2}{c|}{$(2,2)$} \\ \hline
                              & $\eta$      & edges      & $\eta$      & edges      & $\eta$      & edges      \\ \hline
 $A$                  & 0.05        &       248         & 0.05       &         2500      & 0.05       &          6238      \\ \hline
 $B$                  & 0.05        &        248        & 0.05       &        2500        & 0.005       &        624        \\ \hline
 $C$                  & 0.05        &       248        & 0.005        &        250        & 0.005       &       624         \\ \hline
 $D$                  & 0.101        &       500        & 0.01        &        500        & 0.004       &       500         \\ \hline
\end{tabular}
\end{table}
\subsubsection{DRAGON adapts to the data by dynamically chosen penalty parameters {and improves model inference.}}

DRAGON uses an analytical estimate of the minimum of $R$, Eq. (\ref{LedWol}). Here, we show that this estimate dynamically adapts to sample and feature size, and to edge densities $\eta^{(k,l)}$ across a broad range of simulation settings. 

\begin{figure}[t!]
\centerline{\includegraphics[width=.9\textwidth]{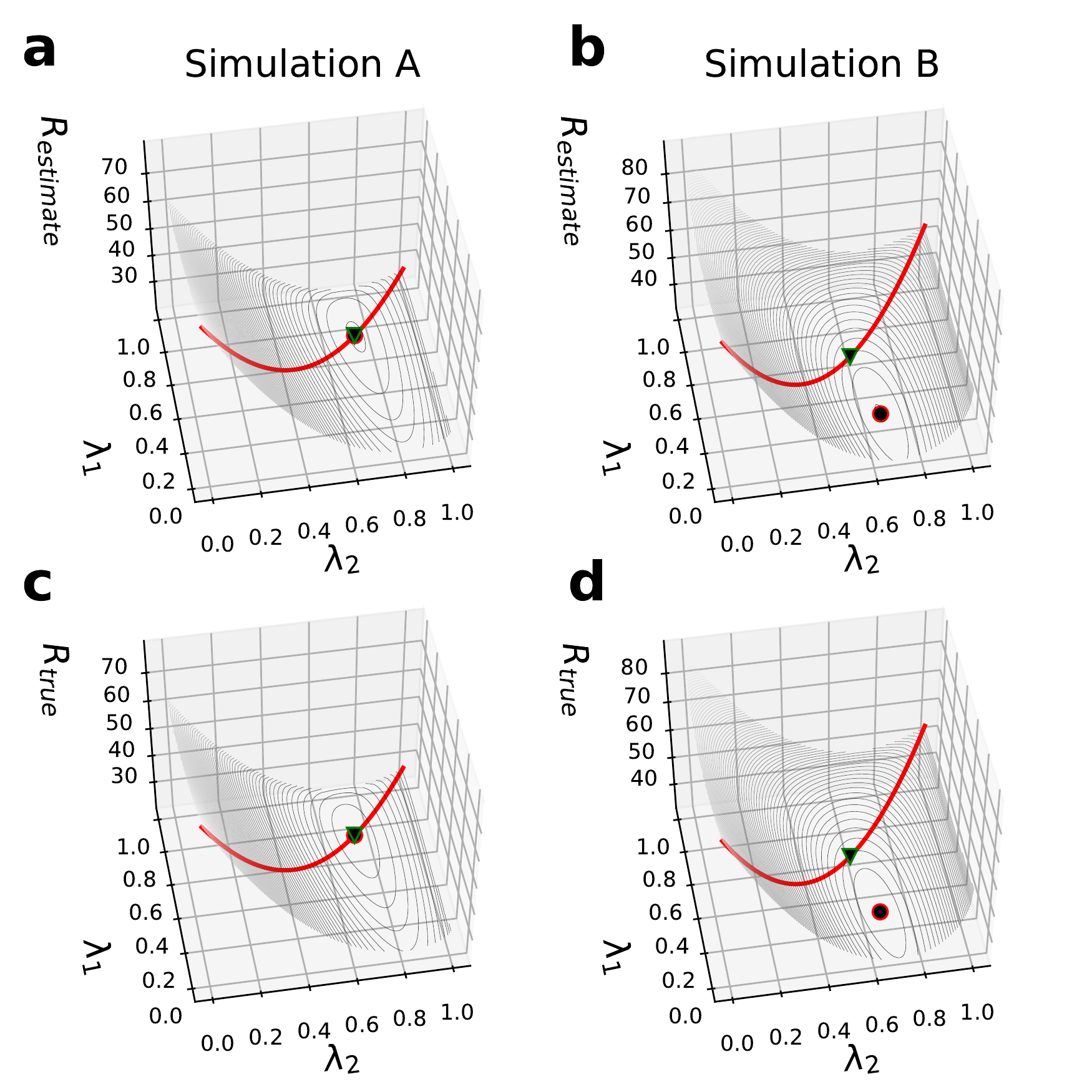}}
\caption{Parameter landscapes for DRAGON. Estimated and true $R$ (upper and lower row) in dependency of $\lambda_1$ and $\lambda_2$ in simulation studies $A$ (left column) and $B$ (right column). Figures a and b show the estimated $R$ for study $A$ and $B$, respectively. Figure c and d show the corresponding ground truth. The red circles indicate the minima for each plot in the $\lambda_1$--$\lambda_2$ plane, and the green triangles give the minima on the diagonal $R$ values shown in red, corresponding to the standard GGM.}
\label{fig_R}
\end{figure}

\textit{Parameter recovery in Simulation A} In simulation study $A$, the two omics layers have equal edge densities $\eta^{(1,1)}=\eta^{(1,2)}=\eta^{(2,2)}=0.05$. Figures \ref{fig_R}a and c show $R$ in dependency of $\lambda_1$ and $\lambda_2$ for the analytical estimate and the ground truth, respectively, for a fixed sample size of $n=5000$. We observed that estimate and ground truth agree remarkably well, and that they provide almost identical estimates for the position of the minimum of the selected penalty parameters, $(\hat\lambda_1,\hat\lambda_2)$. The GGM estimated according to Eq. (\ref{CovShrink}) using the appended data ${X}=[{X}^{(1)},{X}^{(2)}]$ corresponds to the diagonal red lines, such that the two data layers are treated as they would belong to one layer and a single regularization parameter $\lambda_1=\lambda_2=\lambda$ is estimated. We observed that DRAGON correctly estimated $(\hat\lambda_1,\hat\lambda_2)$ to be near to the diagonal for this simulation study. Supplementary Figures S1 and S2 confirm this finding for lower sample sizes, $n=500$ and $n=1000$, respectively. 

\textit{Parameter recovery in Simulation B} In Simulation B, we investigated the influence of the edge densities $\eta^{(k,l)}$ on $R$. For this, we reduced the edge density $\eta^{(2,2)}$ to $\eta^{(2,2)}=0.005$. Again, we observed a remarkable agreement between estimated and true $R$, shown in Figure \ref{fig_R}b and d, respectively, which is also the case for the estimated minima $(\hat\lambda_1,\hat\lambda_2)$. Note, $\hat\lambda_1$ now strongly differs from $\hat\lambda_2$. Since $\lambda_2>\lambda_1$, the second omics data layer is stronger penalized than the first. 

\textit{Parameter recovery in Simulations C and D} We repeated this analysis for two further simulation studies, $C$ and $D$, both of which have unbalanced edge densities with results shown in supplementary Figure S3, S4, and S5 for $n=500$, $n=1,000$, and $n=5,000$, respectively. Here, we also found that DRAGON correctly estimates $R$ and that $\lambda_1$ and $\lambda_2$ are chosen by the algorithm accordingly to minimize $R$. 

{\textit{Model inference in Simulations A-D} We next analyzed how the DRAGON regularization scheme, which augments covariance shrinkage by omics layer-specific regularization parameters, affects model inference.}
We repeated simulation studies $A$ to $D$ 20 times for different sample sizes $n$ {and evaluated the log-likelihood on $n=1,000$ test samples (Figure S6). While we recognize that few omics studies have 1,000 or more samples, single-cell experiments generally assay thousands of cells and large cohort studies are beginning to develop omics databases that include populations of this size or larger.}

We found similar results on an absolute scale (Figures S6a to d). However, results were clearly in favor of DRAGON when we evaluated the log-likelihood difference (Figures S6e to h). This measure has the advantage that it removes variability due to different simulation runs. The green line corresponds to the median log-likelihood difference and the error bands to the $25\%$ and $75\%$ percentiles. Results were slightly in favor of the GGM for the balanced study $A$, but in simuations $B$, $C$, and $D$, DRAGON outperformed the standard GGM as indicated by positive values for the difference. The greatest improvements were seen for the unbalanced simulation $B$. We can also see that as the sample size $n$ increases with the number of predictors $p$ remaining fixed, DRAGON and GGM estimates coincide, as demonstrated by a vanishing log-likelihood difference. 

\subsubsection{DRAGON {outperforms the state of the art with respect to} edge recovery}
Adjusted $p$-values {($q$-values)} were assigned as described in the Materials and Methods section and were used to assess the edge-recovery performance in simulations $A$ to $D$ using receiver operating characteristic (ROC) curves {for DRAGON, GGM, GeneNet, B-NW-SL, D-S-NW-SL, and D-S-GL}. 

\begin{figure}[t!]
\centerline{\includegraphics[width=0.9\textwidth]{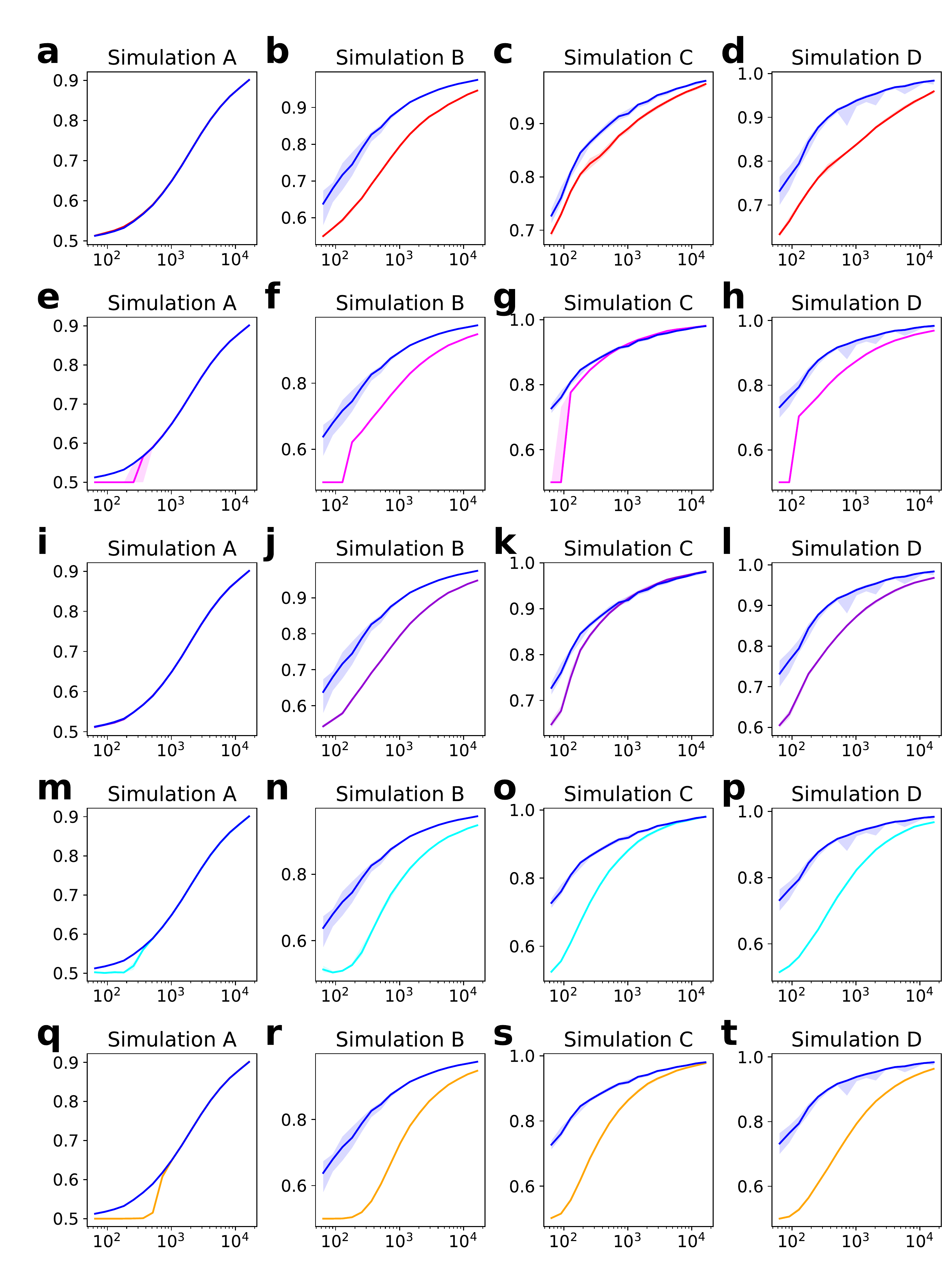}}
\caption{{Area under the ROC curve (AUC) for edge-recovery performance versus sample size $n$ of simulation studies $A$ to $D$ (columns). Each row compares DRAGON (blue) with GGM (red, Fig. a-d), GeneNet (magenta, Fig. e-h), B-NW-SL (purple, Fig. i-l), D-S-NW-SL (cyan, Fig. m-p), and D-S-GL (orange, Fig. q-t), respectively. The lines correspond to the median AUC and the bands to the $25\%$ and $75\%$ percentiles of the distribution.}}
\label{summary}
\end{figure}

\begin{figure}[t!]
\centerline{\includegraphics[width=0.9\textwidth]{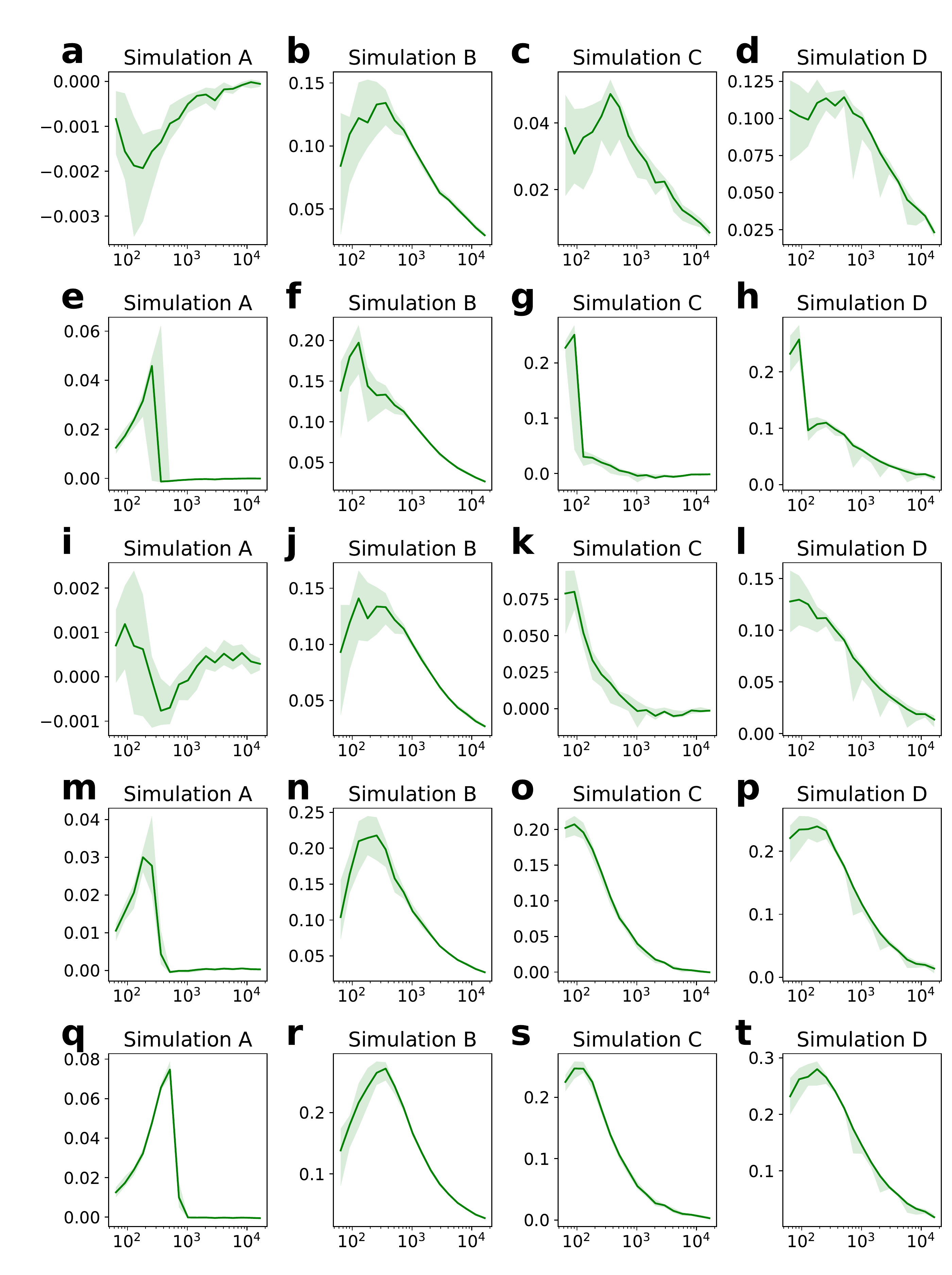}}
\caption{{Area under the ROC curve (AUC) differences versus sample size $n$ of simulation studies $A$ to $D$ (columns). Each row corresponds to the AUC difference DRAGON minus GGM (Fig. a-d), DRAGON minus GeneNet (Fig. e-h), DRAGON minus B-NW-SL (Fig. i-l), DRAGON minus D-S-NW-SL (Fig. m-p), and DRAGON minus D-S-GL (Fig. q-t), respectively. The green lines correspond to the median AUC and the bands to the $25\%$ and $75\%$ percentiles of the distribution.}}
\label{summary_diff}
\end{figure}

{We first compared the results of DRAGON and GGM inference methods.} Figure \ref{summary}a shows how the area under the ROC curve (AUC) depends on the number of training samples $n$ for simulation study $A$. We saw almost identical performance for DRAGON {(blue)} and GGM {(red)} for all sample sizes in our simulation, a result consistent with our previous findings. For simulation studies $B$, $C$, and $D$, the corresponding results are shown in Figures \ref{summary}b to d and illustrate substantial improvements for DRAGON compared to GGM. For example, in simulation study $B$, DRAGON achieved an AUC for $n\approx 200$ samples that is compatible to that of the GGM for $n \approx 1,000$ samples. This improved performance of DRAGON is further illustrated by Figures \ref{summary_diff}{a to d} that shows the respective AUC differences; we found AUC improvements up to $\sim 0.14$ (study $B$), $\sim 0.05$ (study $C$) and $\sim 0.13$ (study $D$). In the balanced simulation scenario $A$ where we expect the GGM and DRAGON to have similar performance, the GGM negligibly outperformed DRAGON, as shown by small negative values in Figure \ref{summary_diff}{a}. As an alternative performance assessment, we used the average area under the precision-recall curve (AUC-PR) {(Suppl. Fig. S7)}. These results also show the performance advantages of  DRAGON compared to GGM {with AUC-PR differences up to $\sim0.07$, $\sim0.03$, $\sim0.08$, for studies $B$ to $D$, respectively.}

{Analogous AUC analyses were performed comparing DRAGON with GeneNet, B-NW-SL, D-S-NW-SL, and D-S-GL with results shown in Figures \ref{summary}e to t and \ref{summary_diff}e to t. Corresponding AUC-PR curves and AUC-PR differences are shown in Suppl. Figures S7 and S8, respectively. For both AUC and AUC-PR, we found in simulated data sets $A$, $B$, and $D$ that DRAGON's improved estimates are of the same size (for GeneNet and B-NW-SL) or substantially larger (for D-S-NW-SL and D-S-GL) than we found in comparing DRAGON to GGM. In simulation study $C$, we found comparable performance of GeneNet and DRAGON across a large range of sample sizes (Fig. \ref{summary}g and \ref{summary_diff}g) but that GeneNet produced an inflated FDR (Suppl. Fig. S9g).} {Table \ref{ConfBounds} summarizes the minimum number of samples $n$ (as well as upper bounds on $p/n$) for each of the simulation studies required to reach network confidence levels AUC$>0.8$. As can be seen, DRAGON reached confidence for substantially smaller $n$ and larger $p/n$ ratios than the other methods.}

\begin{table}
\centering
\caption{\label{ConfBounds}{Lower bounds on $n$ and upper bounds on $p/n$ necessary to achieve confidence in edge recovery defined via an AUC$>0.8$ in simulation studies $A$, $B$, $C$, and $D$. Best performance is indicated in boldface letters.}}
 \scriptsize{
{
\begin{tabular}{|l|l|l|l|l|l|l|l|l|}
\hline
simulation: & \multicolumn{2}{c|}{$A$} & \multicolumn{2}{c|}{$B$} & \multicolumn{2}{c|}{$C$} & \multicolumn{2}{c|}{$D$} \\ \hline
& $n$      & $p/n$      & $n$      & $p/n$      & $n$      & $p/n$ & $n$      & $p/n$      \\ \hline
DRAGON &4096&	0.15&\textbf{256}&\textbf{2.34}&\textbf{128}&\textbf{4.69}&\textbf{181}&\textbf{3.31}\\ \hline
GeneNet &4096&	0.15&1024&0.59&181&	3.31&512	&1.17	\\ \hline
B-NW-SL &4096&	0.15&1024&0.59&181&	3.31&362&1.66\\ \hline
D-S-NW-SL &4096&	0.15&1448&0.41&512&	1.17&1024&0.59\\ \hline
D-S-GL &4096&	0.15&1448&0.41&724&	0.83&1024&0.59\\ \hline
GGM &	4096&	0.15&1024&0.59&181&	3.31&512	&1.17\\ \hline
\end{tabular}
}}
\end{table}

We also verified that DRAGON correctly estimates $p$-values and false discovery rates (FDRs). First, we performed simulations under the null hypothesis of no partial correlation ($\rho=0$) and verified that the $p$-value distributions are flat for different sets of the regularization parameters $\lambda_1$ and $\lambda_2$ (Figures S12 - 13). The simulated data sets were motivated as follows: we recorded the $\lambda_1$ and $\lambda_2$ values for each simulation run in study $A$ to $D$, which yielded the results shown in Figure S12. We extracted the corresponding parameter sets at $n=256$, $n=1,024$ and $n=4,096$ samples, which correspond respectively to a highly regularized estimate, a moderately regularized one, and one with low regularization. This resulted, in total, in 12 pairs $(\lambda_1, \lambda_2)$ with the associated $p$-value distributions under the null hypothesis that are shown in Supplementary Figure S13; as desired, these distributions are flat. Further, we recorded the FDRs corresponding to studies $A$ to $D$ (Supplementary Figures S9a to d and S11) for the empirical estimate of $\kappa$ and the theoretical value $\kappa = n-1-(p-2)$, respectively. As expected, the observed FDR for DRAGON approaches 0.05 as the sample size increases. {Analogous plots for the comparison methods can be found in Supplementary Figure S9e-t. DRAGON generally outperforms the other methods, some of which show inflated FDRs indicating an overly liberal method (GeneNet) or deflated FDRs indicating an overly conservative method (D-S-NW-SL and D-S-GL).}


\subsection{Integrative DRAGON analysis of transcriptome and methylome in TCGA breast cancer specimens}

Epigenetic aberrations such as DNA methylation are associated with altered patterns of gene regulation during the development and progression of complex diseases such as cancer \cite{10.1093/nar/gku1151}. Given the performance advantages of DRAGON over other partial correlation methods, here we present an illustrative application of DRAGON to integrate promoter methylation and gene expression data in 765 breast cancer specimens from The Cancer Genome Atlas (TCGA) \cite{tcga}. We began with a list of 1,590 human transcription factors (TFs) \cite{lambert2018human}, of which 1,557 had annotated promoter methylation data in TCGA, 1,311 had gene expression  data, and 1,297 TFs had both. Below, we outline the analysis of these data; further details regarding the sample population, data preprocessing, model inference, community detection and Reactome pathway enrichment analysis are summarized in Supplementary Sections {4 and 5}. 

\subsubsection{Data acquisition, processing, and estimated DRAGON model}

The GenomicDataCommons \cite{grossman2016toward} R package, version 1.20.1, was used to download TCGA breast cancer data (Project ID TCGA-BRCA; dbGaP Study Accession phs000178) from the Genomic Data Commons. Methylation levels from the Illumina 450k array were processed from raw data into beta values with the TCGA Methylation Array Harmonization Workflow \cite{liftover}, which uses SeSAMe \cite{zhou2018sesame} for signal detection and quality control. Methylation was summarized at the gene level by averaging beta values for probes in the promoter region of each gene of interest and methylation data were transformed to approximate normality with the nonparanormal transformation \cite{liu2009nonparanormal}. RNA-seq data were processed according to the TCGA RNA-Seq Alignment Workflow \cite{tcga_rnaseq} to produce gene expression levels; this pipeline uses the Spliced Transcripts Alignment to a Reference (STAR) \cite{dobin2013star} algorithm to align reads and generate counts which were reported as Transcripts per Million (TPM), among other measures.

\begin{figure}[t!]
    \centering
    \includegraphics[width=0.6\textwidth]{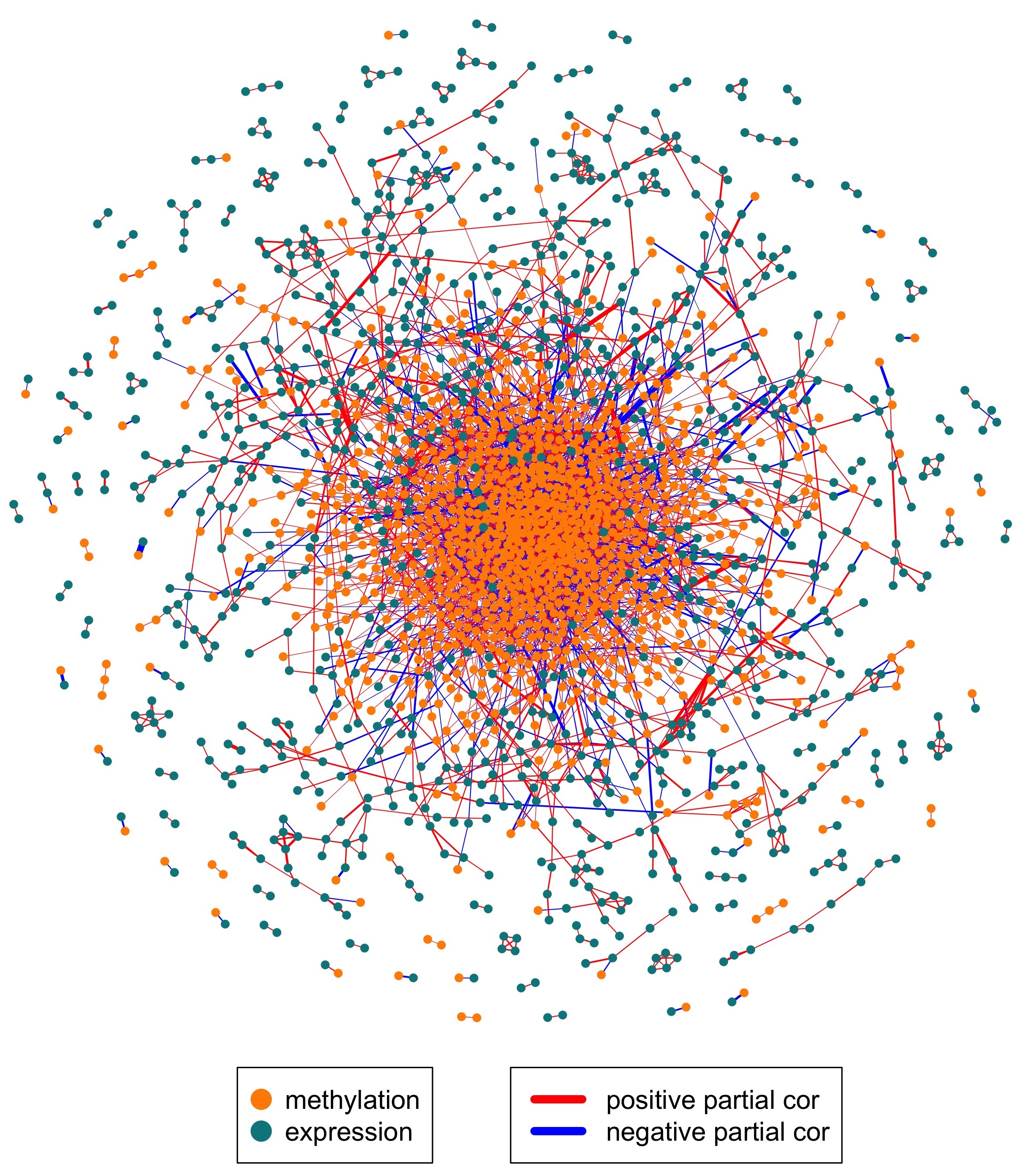}
    \caption{DRAGON multiomic network on TCGA methylation and gene expression data from 795 women with breast cancer.}
    \label{fig:hairball}
\end{figure}

Methylation and expression data for the transcription factors for this breast cancer data set were loaded into DRAGON and a partial correlation network was calculated. A thresholded version of the DRAGON-estimated network consisting of all edges with FDR $< 0.005$ and all nodes with degree $>0$ is shown in Figure \ref{fig:hairball}. The network contains 3,631 edges on 2,106 nodes. 1,168 of the nodes represent methylation (75\% of the methylation variables) and 938 represent gene expression (72 \% of the gene expression variables). 

\subsubsection{DRAGON discovers meaningful promoter methylation-gene expression relationships}

DNA methylation can regulate gene expression by blocking TF binding or by altering the binding of other regulatory proteins \cite{moore2013dna}. In cancer, hypermethylation of tumor suppressor gene promoters can lead to their inactivation by blocking transcription factors and inhibiting the recruitment of the transcriptional machinery \cite{KULIS201027}. On the other hand, DNA methylation results from the actions of proteins (methylases and demethylases) whose levels in turn result from changes in the expression of genes encoding these proteins. DRAGON provides a means to study both changes in gene expression resulting from  methylation and the activation of methylation through changes in gene expression. 

Because of the complex, correlated nature of methylation and gene expression, a multi-omic correlation network based on these data may be too noisy to identify meaningful methylation-expression relationships. To illustrate this, we created a Pearson correlation network on the methylation-gene expression data with {825,256} significant methylation-gene expression edges based on the criteria FDR $< 0.05$. Notably, of these edges, only {862 (0.13\%)} were edges between the expression of a gene and the methylation of its promoter -- the associations we expect functionally to be the most significant ({175 (20\%)} positive and {687 (80\%)} negative edges). In contrast, when using DRAGON to estimate the partial correlation network on the same data, the result was a much sparser network consisting of {769} significant methylation-gene expression edges (FDR $< 0.05$). Of these edges, {333 (43\%)} were associations between expression of a gene and methylation of its promoter ({26 (8\%)} positive and {307 (92\%)} negative partial correlations). The high proportion of negative partial correlations is an important ``sanity check'' on the performance of DRAGON, as promoter methylation suppresses transcription for most genes \cite{KULIS201027}. To further illustrate the discriminatory power of partial correlations relative to simple correlations, we examined the {769} most significant Pearson correlation edges between gene expression and methylation sites (corresponding to the number of gene expression-methylation edges in the DRAGON network) and found only {33 (4\%)} to be associations between expression of a gene and methylation of its promoter. Note that neither the Pearson correlation-based analysis nor the DRAGON analysis use any prior knowledge about methylation site - gene correspondence. 

\begin{figure}
    \centering
    \includegraphics[width=0.48\textwidth]{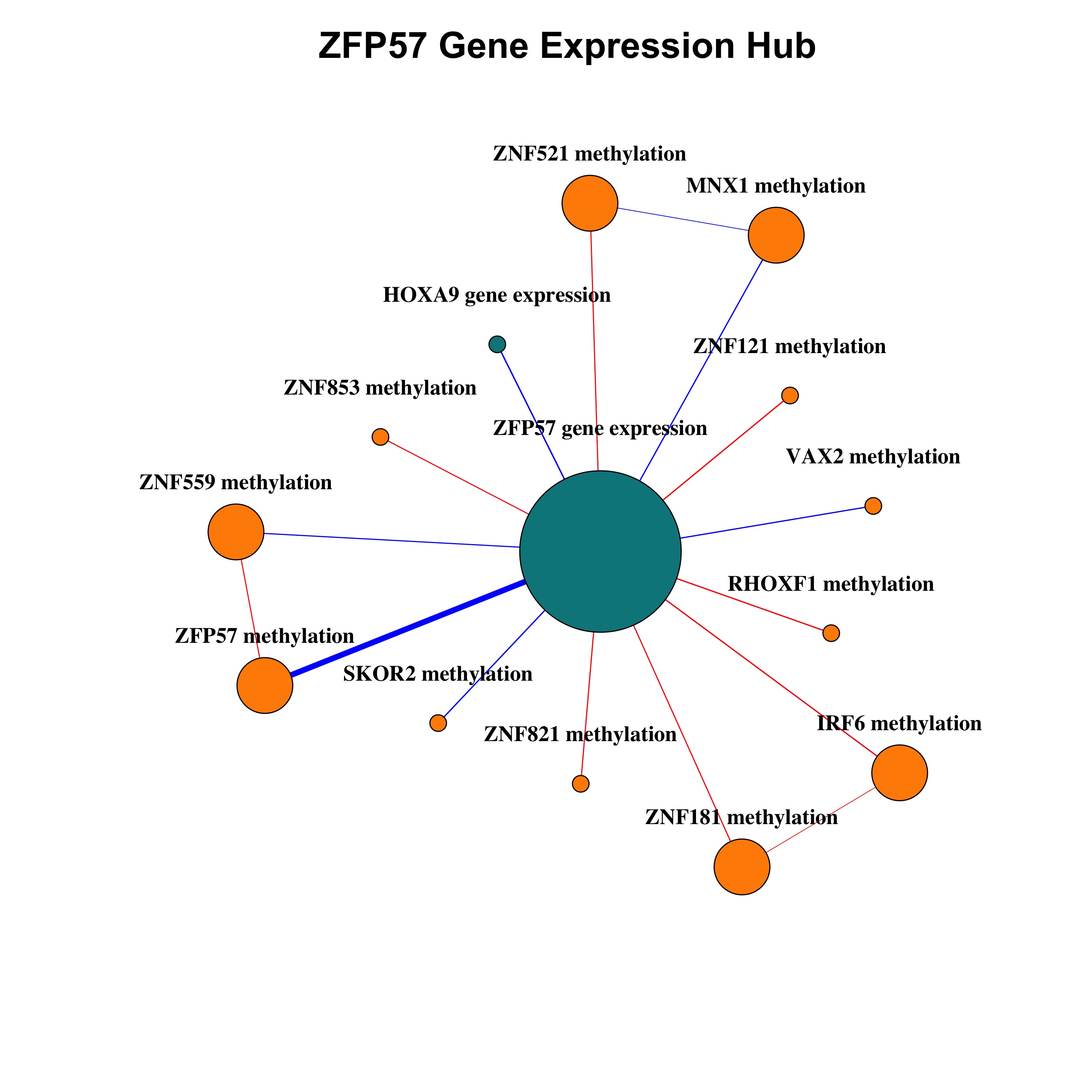}
    \includegraphics[width=0.48\textwidth]{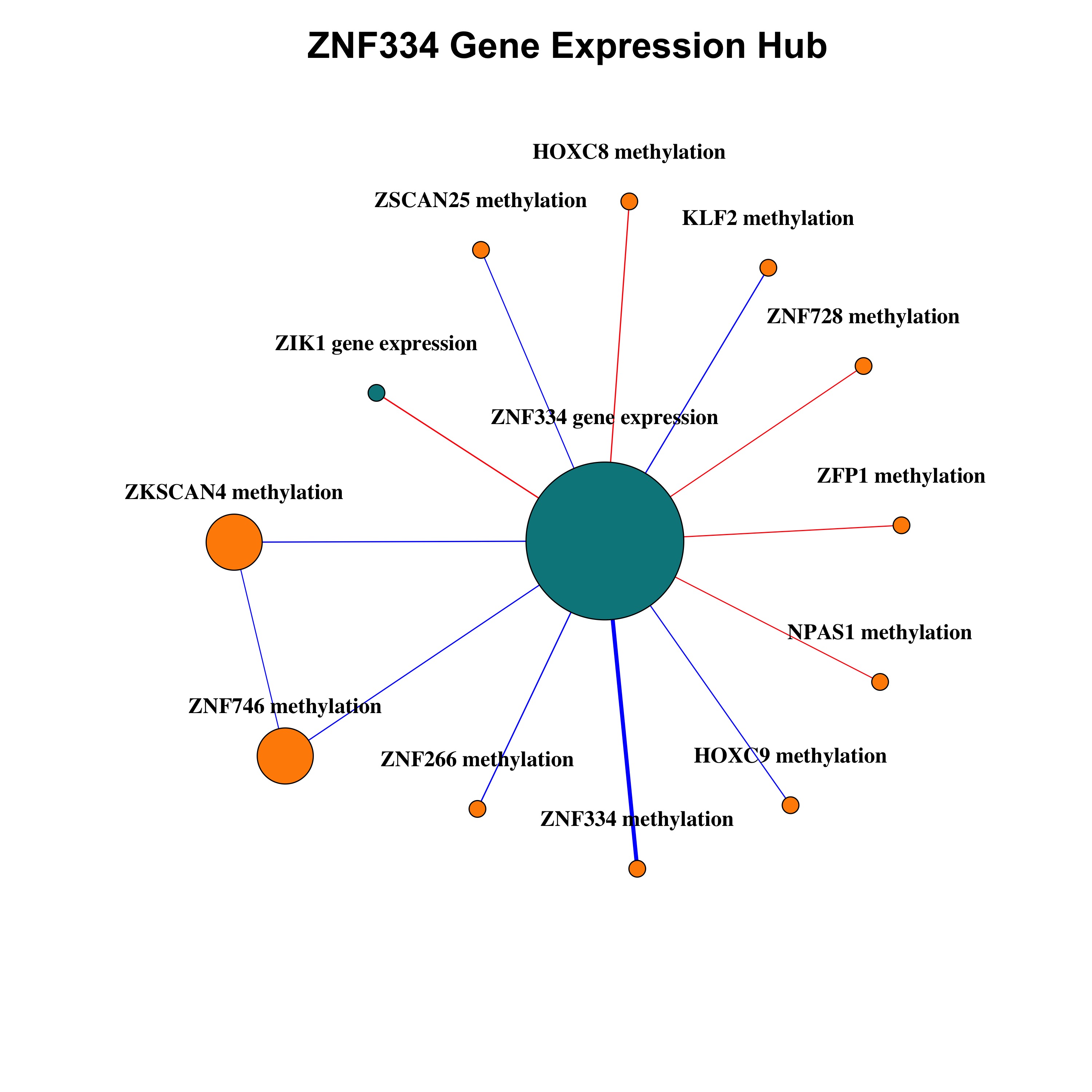} \\
    \caption{{Neighborhoods of ZFP57 and ZNF334, two of the transcription factor gene expression nodes that serve as hubs in the DRAGON breast cancer network. Turquoise nodes represent gene expression and orange nodes represent promoter methylation. Red edges indicate positive partial correlations; blue, negative. Edge width is proportional to the strength of partial correlation. Larger nodes indicate higher node degrees. Edges with FDR $<0.05$ are shown.}}
    \label{fig:hubs}
\end{figure}

Returning to the DRAGON network, we ranked gene expression nodes according to the number of significant edges to methylation nodes they possessed (Table \ref{ranking_GE_methyl}). The top-ranking gene, with 12 edges to methylation nodes (7 positive and 5 negative partial correlations), was ZFP57 (Fig. \ref{fig:hubs}) which is a zinc finger transcription factor containing a KRAB domain and which may play a negative regulatory role \cite{hirasawa2008krab}. The strongest gene expression - promoter methylation edge was observed between ZFP57 and methylation of its own promoter (FDR $<10^{-100}$, $\rho=-0.25$), followed by methylation of the SKOR2 promoter (FDR $<10^{-3}$, $\rho=-0.06$), and methylation of the ZNF121 promoter (FDR $<10^{-3}$, $\rho=0.06$). A comprehensive list of edges is shown in Table \ref{ranking_GE_methyl}. ZFP57 is known to contribute in maintaining the methylation memory of parental origin \cite{shi2019zfp57}. It controls DNA methylation during the earliest multicellular developmental stages at multiple imprinting control regions \cite{mackay2008hypomethylation,takahashi2019znf445}, which is in line with its multiple related methylation sites suggested by DRAGON. Most importantly, ZFP57 has been shown to suppress proliferation of breast cancer cells through down-regulation of the MEST-mediated Wnt/$\beta$-catenin signalling pathway \cite{chen2019zfp57}.

The second TF was ZNF334, another zinc finger protein transcription factor, which had 11 edges to methylation variables (4 positive, 7 negative; Fig. \ref{fig:hubs}). Again, the strongest edge was observed to its own methylation site (FDR $<10^{-34}$, $\rho=-0.1754$), followed by ZNF266 (FDR $=3.6 \cdot 10^{-4} $,  $\rho=-0.058$) and KLF2  (FDR $=7.6 \cdot 10^{-4} $,  $\rho=-0.056$). ZNF334 was recently identified as tumor suppressor of triple-negative breast cancer and higher ZNF334 expression was shown to be associated with better survival outcomes \cite{cheng2022disruption}. DRAGON suggests that this suppression may be due to hypermethylation of its promoter region. A similar pattern has been observed in some other cancers, including hepatocellular carcinoma \cite{sun2022dna}.

The third and fourth top hits were the transcription factors NR6A1 (also known as GCNF) and MYRFL, both showing 10 edges to methylation variables (Supplementary Figure S16). Unlike the other transcription factors we identified, NR6A1 was not related to its own methylation site (FDR $\sim 1$,  $\rho\sim 0$). Its strongest methylation-gene expression edges were observed to methylation sites attributed to the genes FOXO6 (FDR $< 1\cdot 10^{-4}$,  $\rho= 0.061$) and IKZF3 (FDR $< 1\cdot 10^{-4}$,  $\rho= 0.060$). FOXO6 is a transcription factor known to play multiple roles in breast cancer. Its downregulation is implicated in promotion of the epithelial–mesenchymal transition, in migration and proliferation of breast cancer cells, and in reduced cell resistance to the anti-cancer drug paclitaxel through the PI3K/Akt signaling pathway \cite{ye2018downregulation, yu2020knockdown}. IKZF3 is a member of the Ikaros family of zinc-finger proteins that has been shown to work with other transcription factors to regulate immune response in breast cancer \cite{da2017transcription} and its knockdown has been shown to dramatically increase breast cancer response to chimeric antigen receptor T-cell (CAR-T) therapy \cite{zou2022ikzf3}.

In \textit{mus musculus}, NR6A1 has been shown to interact with DNMT3B (DNA (cytosine-5)-methyltransferase 3B) to induce promoter methylation of the genes  Oct-3/4  \cite{sato2006orphan}. DNMT3B together with DNMT3A is essential for the \textit{de novo} methylation in early development \cite{okano1999dna}. In addition to its role as a transcription factor, NRA61 is an orphan nuclear receptor normally expressed in germ cells of gonads and highly expressed in triple-negative and ER + HER2 - breast cancer and so has been suggested as an ideal drug target \cite{willis2015enriched}. MYRFL follows the more typical pattern, having its strongest methylation-gene expression edge to its own methylation site (FDR $< 1\cdot 10^{-7} $,  $\rho=-0.073$). MYRFL encodes a transcription factor that is required for central nervous system myelination and it has been identified as a member of a regulatory cluster of genes on chromosome 12 that has been associated with elevated risk of breast cancer \cite{madsen2018reparameterization}.

Although this analysis does not fully account for either the complexities of breast cancer and its subtypes or for the interplay of regulatory mechanisms active in cells and limited exploration to the transcription factors themselves, it already paints a compelling picture of the interplay between epigenetic regulation through altered patterns of methylation in breast cancer and activation or repression of key regulatory proteins that control breast cancer risk, cell proliferation, and response to various therapeutic interventions. 

\begin{table*}[ht]
\caption{\label{ranking_GE_methyl}{Summary of the 4 gene-expression variables with most edges to methylation variables ordered by rank. Methylation variables (last column) are ordered by partial correlations (from high to low absolute values), where the $+/-$ signs in brackets give the signs of the estimated partial correlations.}}
{
\centering
\footnotesize{
\begin{tabular}{rlll}
  \hline
rank & TF RNA & \#GE-methyl. edges & methylation variables \\ 
  \hline
1 & ZFP57 & 12 & ZFP57(-), SKOR2(-), ZNF121(+), IRF6(+), RHOXF1(+), MNX1(-), VAX2(-),\\ &&&ZNF821(+), ZNF181(+), ZNF559(-), ZNF521(+), ZNF853(+) \\ 
\hline
  2 & ZNF334 & 11 & ZNF334(-), ZNF266(-), KLF2(-), ZKSCAN4(-), HOXC8(+), ZNF746(-),\\
&&& HOXC9(-), ZFP1(+), ZNF728(+), ZSCAN25(-), NPAS1(+) \\\hline 
  3 & NR6A1 & 10 & FOXO6(+), IKZF3(+), JRK(-), NR5A1(+), NFIC(+), ZNF654(-), FOXR1(-),\\
&&&
ETV2(+), ELK3(+), FOXP1(-) \\\hline
  4 & MYRFL & 10 & MYRFL(-), ZFP69B(-), AHR(-), HOXB7(+), ZNF467(-), THAP12(+), ZNF251(-), \\
  &&&MTF1(+), ARID5A(+), RORB(+) \\ \hline
\end{tabular}}}
\end{table*}

\subsubsection{Community detection and enrichment analysis}

To further analyze the complex structure of the estimated network, we clustered nodes using community detection on the DRAGON-estimated network thresholded at FDR $< 0.005$. This threshold was based on inspection of the distribution of FDR-corrected $p$-values; Supplementary Figure S17. Community detection was performed using the \texttt{cluster\_fast\_greedy} algorithm as implemented in the \texttt{igraph} R package \cite{pons2005computing,csardi2006igraph} (see Supplement). Using this algorithm, 169 communities were detected, 59 of which had at least 5 nodes. 

To assess the enrichment of methylation-gene expression communities for functions potentially related to breast cancer, we performed an over-representation analysis (ORA) for Reactome gene sets \cite{gillespie2022reactome} within each of the 59 communities with at least 5 nodes (for details of the ORA, see Supplement). The \texttt{fora} function of the \texttt{fgsea} R package was used to conduct the ORA \cite{korotkevich2021fast}. Reactome gene sets with at least 3 genes were considered (\texttt{minSize} = 3). For the analysis presented here, it should be noted that each transcription factor gene may appear twice, once based on its expression and once based on its methylation. Therefore, the universe of possible genes considered for the ORA (and used to set background expectations) is twice the size of the number of TFs included in the DRAGON model. We also performed an ORA assessing enrichment for methylation only and one assessing enrichment for expression only. A Reactome pathway was identified as over-represented in a community if its FDR was less than 0.05 in at least one of these three ORAs (Benjamini-Hochberg FDR as implemented in \texttt{fgsea}). Ten of the 59 communities of size $\geq$ 5 genes were enriched for at least one REACTOME pathway at FDR $< 0.05$; complete results of the enrichment analysis are available in Supplementary Data 1. Here, we highlight two communities that illustrate DRAGON's ability to provide unexpected insight into disease processes. 

\begin{figure}
    \centering
    \includegraphics[width=0.49\textwidth]{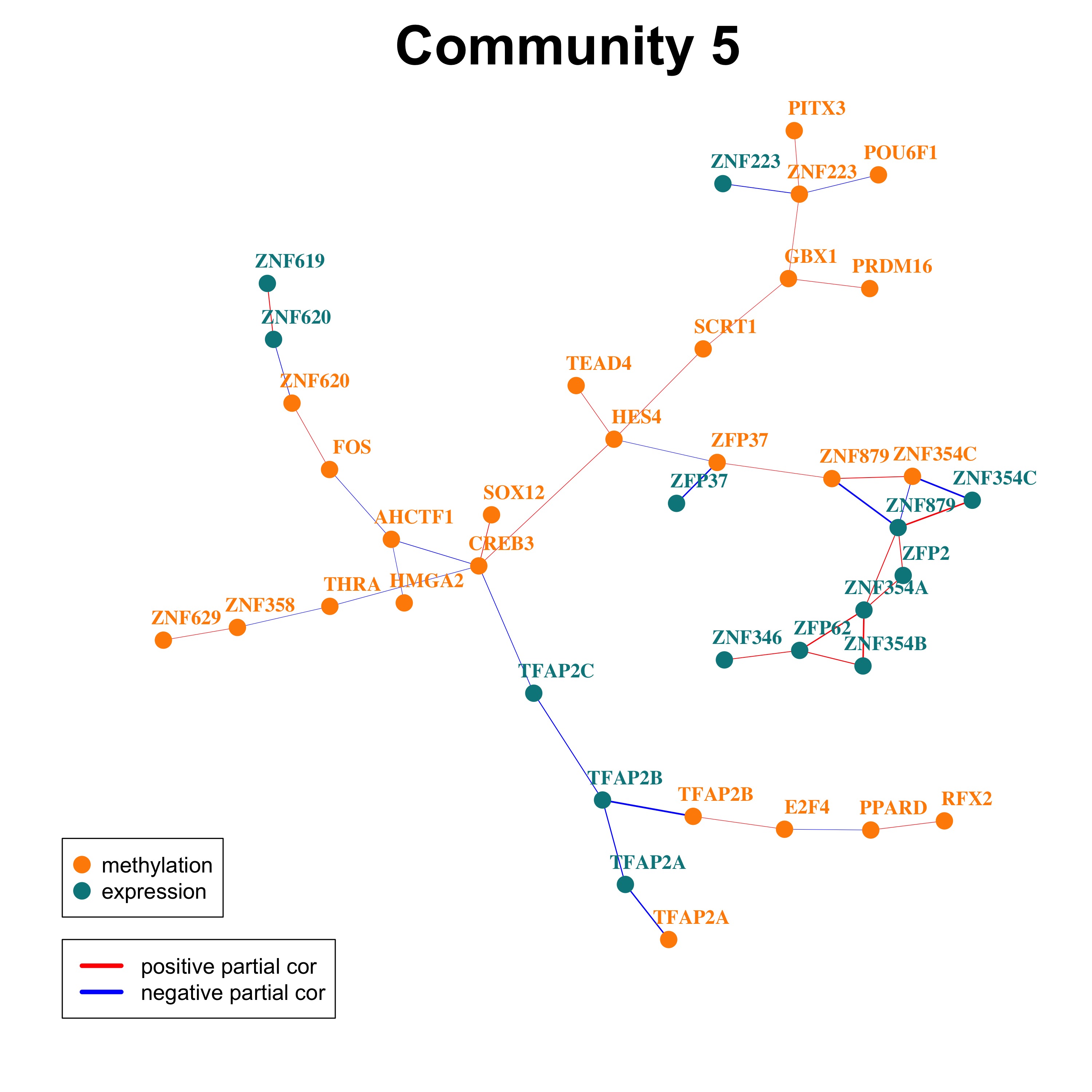}
    \includegraphics[width=0.49\textwidth]{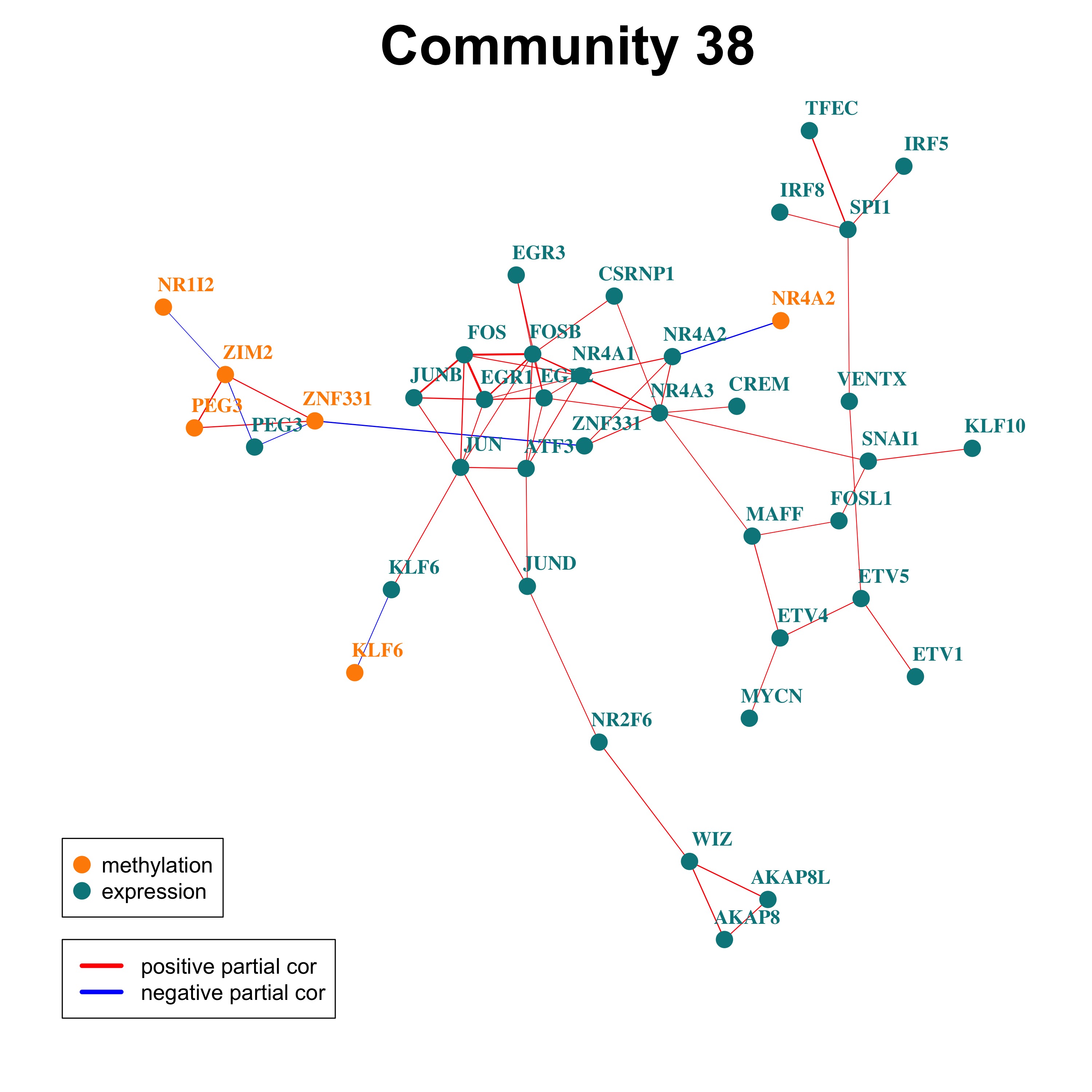}
    \caption{{Example communities of interest in the DRAGON breast cancer gene expression-methylation network. Community 5 contains the TFAP2 family of transcription factors. Community 38 contains several hallmark cancer genes that are highly connected.}}
    \label{fig:DRAGON_comms}
\end{figure}

Community 5 consists of 39 TF nodes, 14 based on TF gene expression and 25 based on TF promoter methylation (Figure \ref{fig:DRAGON_comms}). The TF set in community 5 is enriched for several Reactome pathways related to the TFAP2 family of transcription factors (Supplementary Data 1). TFAP2C has been implicated in estrogen response signaling in breast cancer, which plays a major role in breast cancer development, progression, and therapeutic response\cite{woodfield2007tfap2c}; estrogen response signaling is also a key determinant of breast cancer molecular subtype \cite{eroles2012molecular}. 

To explore the role of the TFAP2 family in our DRAGON network, we obtained mRNA-based subtype classifications (basal, HER2, luminal A, luminal B, and normal) for the tumor samples using the \texttt{PanCancerAtlas\_subtypes} function from the \texttt{TCGAbiolinks} R package \cite{colaprico2016tcgabiolinks}. We then evaluated subtype-specific methylation and expression among the TFAP2 genes represented in Community 5 (TFAP2A methylation and expression, TFAP2B methylation and expression, TFAP2C expression; Supplementary Figure S18).

TFAP2A and TFAP2B methylation both differed significantly based on subtype (Kruskal-Wallis test, TFAP2A $p = 3.3\cdot10^{-4}$; TFAP2B $p < 1\cdot 10^{-15}$) as did TFAP2A, TFAP2B, and TFAP2C expression (TFAP2A $p = 1.9\cdot 10^{-12}$, TFAP2B $p < 1\cdot 10^{-15}$, TFAP2C $p <1.9\cdot 10^{-10}$). However, TFAP2C methylation, which was notably excluded from this community, did not differ significantly between subtypes ($p =0.17$). To illustrate the multi-omic phenotyping possible with DRAGON communities, we investigated TFAP2B further. The median TFAP2B promoter methylation beta level was 0.55 in basal samples versus 0.17 in non-basal samples; the median TFAP2B expression among basal samples was 0.49 TPM vs. a median of 36.46 TPM in non-basal samples. This multi-omic phenotype of increased methylation and decreased expression in basal samples follows the canonical paradigm that promoter methylation results in gene silencing. To explore the predictive power of this joint information, we constructed a logistic regression model for multi-omic subtype classification, regressing basal vs. non-basal subtype against TFAP2B methylation, expression, and an interaction term between the two. In the resulting model, TFAP2B methylation and the interaction term between TFAP2B methylation and expression were both significant (methylation: OR = 147.76, 95 \% CI: [44.21, 493.82]; methylation*expression: OR = 0.873, 95 \% CI: [0.797, 0.957]) while TFAP2B expression was not (OR = 1.004, 95 \% CI: [0.989, 1.018]). The AUROC of this model was 0.85; in contrast, the AUROC of a similar model using TFAP2B methylation alone as a predictor was 0.82 and using TFAP2B expression alone, 0.79, demonstrating the synergistic power of multi-omic features for class discrimination. Although this classification model does not outperform the expression-based subtype classification used by the TCGA, these results are nonetheless impressive given that they are based on two measurements of the omic state of a single gene.
 
Community 38 consists of 40 nodes, six of which represent promoter methylation and 34 of which represent gene expression (Figure \ref{fig:DRAGON_comms}). This community is enriched for 21 Reactome pathways, including signaling by receptor tyrosine kinases (FDR = $4.7 \cdot 10^{-9}$), estrogen-dependent gene expression (FDR = $1.5 \cdot 10^{-3}$), PTEN regulation (FDR = 0.02), and MAPK family signaling cascades (FDR = 0.02). The nodes primarily driving these enrichments include EGR1, EGR2, EGR3, FOS, FOSB, JUNB, and JUND. These seven nodes comprise a tightly co-expressed subgraph of Community 38 and are well-known players in cancer signaling \cite{lee2001cross,wang2021role}. 

NR4A2 methylation and KLF6 methylation are both degree-one nodes in Community 38; their only edges indicate negative partial correlation with their own expression. It may be that differential methylation of NR4A2 and KLF6 drives subtype-specific activity in the pathways of this community by modulating the expression of NR4A2 and KLF6, which are connected via FOS and JUN expression to the eight-node cluster of coexpressed TFs. In comparing these nodes between subtypes, we found that both KLF6 and NR4A2 are significantly differentially methylated (KLF6: Kruskal-Wallis chi-squared = 93.04, df = 4, $p$ $<2.2\cdot 10^{-16}$; NR4A2: Kruskal-Wallis chi-squared = 17.98, df = 4, $p$ = 0.001; Supplementary Figure S21). To provide a benchmark for differential methylation between subtypes, we performed the same statistical test for a ``housekeeping'' transcription factor (ATF1), which showed no significant difference in methylation (Kruskal-Wallis chi-squared = 6.62, df = 4, $p$ = 0.158; Supplementary Figure S19), indicating that the differential methylation of these key transcription factors may play a substantial role in breast cancer beyond those already discussed, particularly given their association with the EGR/FOS/JUN expression cluster and the importance of these transcription factors in a wide range of cancer processes.

\section{Conclusion and Summary}
{Regulation of transcriptional processes in the cell involve multiple interacting partners that include transcription factors and their expression, regulation of their targets, and epigenetic factors such as DNA methylation that may enhance or disrupt regulatory interactions. Simple measures such as correlation fail to capture meaningful regulatory associations and can be dominated by spurious correlations between genes that are expressed at relatively low levels or that exhibit similar patterns of expression due to factors unrelated to the biological state of the system. Partial correlation analysis allows better discrimination between potentially causal associations between regulators and their regulatory targets and may lead to greater insight into the underlying biology of the systems we choose to study. The problem in conducting such an analysis is that different types of omics data often present with different scales, biases, and error distributions.}

DRAGON is a partial correlation framework optimized for the integration of multiple ``layers'' of omics data into a unified association network that allows us to understand both associations between biological variables such as gene expression and the potential drivers of the observed correlations. DRAGON is based on Gaussian Graphical Models (GGMs) and uses a regularization scheme to optimize the trade-off between the network's complexity and its estimation accuracy while explicitly taking into account the distinct data characteristics of the various omics data types used in the model. DRAGON accounts for differences in edge densities and feature sizes, enabling improved estimation of partial correlations compared to layer-agnostic GGMs. The advantages of DRAGON are  particularly evident in simulations when the number of variables $p$ is the same order of magnitude or exceeds the sample size $n$, as is the case in nearly all omics experiments.

We recognize that DRAGON has some limitations. DRAGON's GGM framework assumes multivariate normally distributed data which, for biological data, generally does not hold. For continuous distributions, data transformations such as the nonparanormal transformation \cite{liu2009nonparanormal} can be used to adjust input data to be approximately normally distributed; this approach, for example, allowed us to use DRAGON with methylation and gene expression data. However, other omics data types such as single-nucleotide polymorphism (SNP) data are categorical or ordinal, and alternative methods are needed to build these important regulatory elements into the DRAGON framework. Extending DRAGON to Mixed Graphical Models \cite{lee2015learning, altenbuchinger2019multi}, which incorporate both continuous and discrete variables, could allow us to overcome this limitation of the current implementation. 

{It is also worth noting that DRAGON does not incorporate pre-defined network structures. Structured probabilistic graphical models have been studied fairly extensively \cite{ambroise2009inferring,ma2016network,siahpirani2017prior,zhuang2022augmented} as they allow users to bias networks towards a given structure consistent with biological prior knowledge. Work by our group and others has demonstrated the power of introducing soft, knowledge-based constraints into optimization problems such as gene regulatory network inference \cite{wolpert1997no,glass2013passing}. In the case of DRAGON, this could be achieved by \textit{a priori} removing edges from the model based on known regulatory relationships by, for example, estimating the inverse biased covariance matrix with \textit{a priori} specified zero entries \cite{hastie2009elements}. Alternatively, one could construct a model in which ``established'' associations between elements are more likely to be included into the network by shifting their weights at initiation \cite{ma2016network} or through the use of modified target matrices $T^{(k)}$ in Eq. (\ref{DRAGON}). Both approaches, alone or in combination, would need to be carefully tested taking into account the effect that a bias on the network structure has on estimates of significance levels.} 

{Although DRAGON represents an important step forward given our ability to collect increasingly large, multi-omic data sets, it is important to recognize that many problems in network inference are not addressed by DRAGON. Partial correlation relies on the data we have available, and many regulatory data types simply cannot be collected simultaneously and so remain hidden. Such hidden variables can lead to spurious associations and hamper the interpretation of networks in general. New technologies may provide additional omics layers, but integration of these will require additional methodological advances and the development of robust and scalable computational models.
}

{
Second, conditional independence is only one way to model associations in biological data. Conditional independence relationships encode the factorization properties of probability distributions and while appealing as a model, it is difficult to state definitively how this concept maps to the complexity of regulatory processes in biological systems. However, the same limitation holds for all measures of ``relatedness,'' comprising mutual exclusivity \cite{margolin2006aracne} and multivariate information measures \cite{chan2017gene}. Elucidation of which type of similarity measure is most appropriate for inference of networks from biological data remains an open challenge \cite{marbach2012wisdom}, but probabilistic graphical models have been shown to perform well relative to other approaches (see the lasso regression methods in \cite{marbach2012wisdom} as an example). To give a clear answer, experimental data together with the corresponding ground truth is optimally needed, but our use of inference methods underlies our present inability to determine the true regulatory processes that drive biological states. Consequently, method comparison of probabilistic graphical models that allows us to specify appropriate benchmarks with respect to edge recovery, maximum likelihood, and false discovery rate are the best available methods for testing the basic process of network inference and modeling, and assessment of biological insight gained trough model analysis remains a key element of validating new methods. On both accounts, DRAGON performs well giving limited data sets that reflect those typically available in omics studies of human health and disease.  
}
{
We also note that our analysis focused on observational data only, and the inferred networks are undirected. Statements on causality from observational data are difficult given correlation-based models, although it is possible to provide lower bounds on causal effects \cite{maathuis2010predicting}. However, our biological understanding of cause and effect can guide us, at least in part. Nevertheless, additional work needs to be done to address causality in the context of multi-omics and the use of structured approaches that incorporate prior causal knowledge may be an important next step. A simple adherence to the ``central dogma of molecular biology,'' that DNA makes RNA, and RNA makes protein, could assist in defining causal relationships. Importantly, specifying a prior belief should not prohibit the inference of more complex regulatory mechanisms contradicting these beliefs: otherwise, we stand to discover only what we already know.}

DRAGON represents a significant contribution to network inference by presenting a framework for modeling of partial correlations across multiple layers of omics data. DRAGON-estimated networks provide new insights into regulatory processes that may be overlooked by other methods as they are capable of identifying direct multi-omic relationships via a Gaussian graphical model framework. DRAGON is easy to use and freely available as open source software in the Network Zoo package (netZooPy v0.8; netzoo.github.io).

\section{Code and data availability}
DRAGON is available through the Network Zoo package (netZooPy v0.8; netzoo.github.io) and a vignette to demonstrate its use can be found in netBooks (netbooks.networkmedicine.org). Code to reproduce the TCGA breast cancer methylation analysis is available on Github at \url{https://github.com/katehoffshutta/DRAGON-TCGA-BRCA}.

\section{Support}
KHS was supported by grant 2T32HL007427 from the National Heart, Lung, and Blood Institute of the US National Institutes of Health. DW, RB, MBG, JQ, and MA were supported by grants from the US National Cancer Institute (1R35CA220523 and 1U24CA231846). DLD was supported by grants from the US National Institutes of Health, award numbers R01HG011393 and P01114501. The work of HUZ and MA was supported by the German Federal Ministry of Education and Research (BMBF) within the framework of the e:Med research and funding concept (grant no. 01ZX1912A and 01ZX1912C). 

{
\section{Acknowledgments}
The results shown here are in whole or part based upon data generated by the TCGA Research Network: https://www.cancer.gov/tcga.}

\bibliographystyle{unsrt}

\begin{thebibliography}{10}

\bibitem{stelzl2005human}
Ulrich Stelzl, Uwe Worm, Maciej Lalowski, Christian Haenig, Felix~H Brembeck,
  Heike Goehler, Martin Stroedicke, Martina Zenkner, Anke Schoenherr, Susanne
  Koeppen, et~al.
\newblock A human protein-protein interaction network: a resource for
  annotating the proteome.
\newblock {\em Cell}, 122(6):957--968, 2005.

\bibitem{rual2005towards}
Jean-Fran{\c{c}}ois Rual, Kavitha Venkatesan, Tong Hao, Tomoko
  Hirozane-Kishikawa, Am{\'e}lie Dricot, Ning Li, Gabriel~F Berriz, Francis~D
  Gibbons, Matija Dreze, Nono Ayivi-Guedehoussou, et~al.
\newblock Towards a proteome-scale map of the human protein--protein
  interaction network.
\newblock {\em Nature}, 437(7062):1173--1178, 2005.

\bibitem{szklarczyk2019string}
Damian Szklarczyk, Annika~L Gable, David Lyon, Alexander Junge, Stefan Wyder,
  Jaime Huerta-Cepas, Milan Simonovic, Nadezhda~T Doncheva, John~H Morris, Peer
  Bork, et~al.
\newblock String v11: protein--protein association networks with increased
  coverage, supporting functional discovery in genome-wide experimental
  datasets.
\newblock {\em Nucleic acids research}, 47(D1):D607--D613, 2019.

\bibitem{markowetz2007inferring}
Florian Markowetz and Rainer Spang.
\newblock Inferring cellular networks--a review.
\newblock {\em BMC bioinformatics}, 8(6):S5, 2007.

\bibitem{glass2013passing}
Kimberly Glass, Curtis Huttenhower, John Quackenbush, and Guo-Cheng Yuan.
\newblock Passing messages between biological networks to refine predicted
  interactions.
\newblock {\em PloS one}, 8(5), 2013.

\bibitem{aoki2007approaches}
Koh Aoki, Yoshiyuki Ogata, and Daisuke Shibata.
\newblock Approaches for extracting practical information from gene
  co-expression networks in plant biology.
\newblock {\em Plant and Cell Physiology}, 48(3):381--390, 2007.

\bibitem{altenbuchinger2020gaussian}
Michael Altenbuchinger, Antoine Weihs, John Quackenbush, Hans~J{\"o}rgen Grabe,
  and Helena~U Zacharias.
\newblock Gaussian and mixed graphical models as (multi-) omics data analysis
  tools.
\newblock {\em Biochimica et Biophysica Acta (BBA)-Gene Regulatory Mechanisms},
  1863(6):194418, 2020.

\bibitem{aldrich1995correlations}
John Aldrich et~al.
\newblock Correlations genuine and spurious in pearson and yule.
\newblock {\em Statistical science}, 10(4):364--376, 1995.

\bibitem{margolin2006aracne}
Adam~A Margolin, Ilya Nemenman, Katia Basso, Chris Wiggins, Gustavo
  Stolovitzky, Riccardo Dalla~Favera, and Andrea Califano.
\newblock Aracne: an algorithm for the reconstruction of gene regulatory
  networks in a mammalian cellular context.
\newblock In {\em BMC bioinformatics}, volume~7, pages 1--15. BioMed Central,
  2006.

\bibitem{butte1999mutual}
Atul~J Butte and Isaac~S Kohane.
\newblock Mutual information relevance networks: functional genomic clustering
  using pairwise entropy measurements.
\newblock In {\em Biocomputing 2000}, pages 418--429. World Scientific, 1999.

\bibitem{wille2004sparse}
Anja Wille, Philip Zimmermann, Eva Vranov{\'a}, Andreas F{\"u}rholz, Oliver
  Laule, Stefan Bleuler, Lars Hennig, Amela Preli{\'c}, Peter von Rohr, Lothar
  Thiele, et~al.
\newblock Sparse graphical gaussian modeling of the isoprenoid gene network in
  arabidopsis thaliana.
\newblock {\em Genome biology}, 5(11):R92, 2004.

\bibitem{schafer2005shrinkage}
Juliane Sch{\"a}fer and Korbinian Strimmer.
\newblock A shrinkage approach to large-scale covariance matrix estimation and
  implications for functional genomics.
\newblock {\em Statistical applications in genetics and molecular biology},
  4(1), 2005.

\bibitem{krumsiek2011gaussian}
Jan Krumsiek, Karsten Suhre, Thomas Illig, Jerzy Adamski, and Fabian~J Theis.
\newblock Gaussian graphical modeling reconstructs pathway reactions from
  high-throughput metabolomics data.
\newblock {\em BMC systems biology}, 5(1):21, 2011.

\bibitem{ghanbari2019distance}
Mahsa Ghanbari, Julia Lasserre, and Martin Vingron.
\newblock The distance precision matrix: computing networks from non-linear
  relationships.
\newblock {\em Bioinformatics}, 35(6):1009--1017, 2019.

\bibitem{cao2018joint}
Junyue Cao, Darren~A Cusanovich, Vijay Ramani, Delasa Aghamirzaie, Hannah~A
  Pliner, Andrew~J Hill, Riza~M Daza, Jose~L McFaline-Figueroa, Jonathan~S
  Packer, Lena Christiansen, et~al.
\newblock Joint profiling of chromatin accessibility and gene expression in
  thousands of single cells.
\newblock {\em Science}, 361(6409):1380--1385, 2018.

\bibitem{meinshausen2006high}
Nicolai Meinshausen, Peter B{\"u}hlmann, et~al.
\newblock High-dimensional graphs and variable selection with the lasso.
\newblock {\em The annals of statistics}, 34(3):1436--1462, 2006.

\bibitem{friedman2008sparse}
Jerome Friedman, Trevor Hastie, and Robert Tibshirani.
\newblock Sparse inverse covariance estimation with the graphical lasso.
\newblock {\em Biostatistics}, 9(3):432--441, 2008.

\bibitem{lauritzen1996graphical}
Steffen~L Lauritzen.
\newblock {\em Graphical models}, volume~17.
\newblock Clarendon Press, 1996.

\bibitem{bishop2006pattern}
Christopher~M Bishop.
\newblock {\em Pattern recognition and machine learning}.
\newblock springer, 2006.

\bibitem{ledoit2000well}
Olivier Ledoit and Michael Wolf.
\newblock A well conditioned estimator for large dimensional covariance
  matrices.
\newblock 2000.

\bibitem{bernal2019exact}
Victor Bernal, Rainer Bischoff, Victor Guryev, Marco Grzegorczyk, and Peter
  Horvatovich.
\newblock Exact hypothesis testing for shrinkage-based gaussian graphical
  models.
\newblock {\em Bioinformatics}, 35(23):5011--5017, 2019.

\bibitem{benjamini1995controlling}
Yoav Benjamini and Yosef Hochberg.
\newblock Controlling the false discovery rate: a practical and powerful
  approach to multiple testing.
\newblock {\em Journal of the Royal statistical society: series B
  (Methodological)}, 57(1):289--300, 1995.

\bibitem{GeneNet}
Juliane Schaefer, Rainer Opgen-Rhein, and Korbinian Strimmer.
\newblock {\em GeneNet: Modeling and Inferring Gene Networks}, 2020.
\newblock R package version 1.2.15.

\bibitem{schafer2005empirical}
Juliane Sch{\"a}fer and Korbinian Strimmer.
\newblock An empirical bayes approach to inferring large-scale gene association
  networks.
\newblock {\em Bioinformatics}, 21(6):754--764, 2005.

\bibitem{ren2015asymptotic}
Zhao Ren, Tingni Sun, Cun-Hui Zhang, and Harrison~H Zhou.
\newblock Asymptotic normality and optimalities in estimation of large gaussian
  graphical models.
\newblock {\em The Annals of Statistics}, 43(3):991--1026, 2015.

\bibitem{zhang2018silggm}
Rong Zhang, Zhao Ren, and Wei Chen.
\newblock Silggm: An extensive r package for efficient statistical inference in
  large-scale gene networks.
\newblock {\em PLoS computational biology}, 14(8):e1006369, 2018.

\bibitem{jankova2017honest}
Jana Jankov{\'a} and Sara van~de Geer.
\newblock Honest confidence regions and optimality in high-dimensional
  precision matrix estimation.
\newblock {\em Test}, 26(1):143--162, 2017.

\bibitem{jankova2015confidence}
Jana Jankova and Sara Van De~Geer.
\newblock Confidence intervals for high-dimensional inverse covariance
  estimation.
\newblock {\em Electronic Journal of Statistics}, 9(1):1205--1229, 2015.

\bibitem{10.1093/nar/gku1151}
Wei-Yun Huang, Sheng-Da Hsu, Hsi-Yuan Huang, Yi-Ming Sun, Chih-Hung Chou,
  Shun-Long Weng, and Hsien-Da Huang.
\newblock {MethHC: a database of DNA methylation and gene expression in human
  cancer}.
\newblock {\em Nucleic Acids Research}, 43(D1):D856--D861, 11 2014.

\bibitem{tcga}
Katarzyna Tomczak, Patrycja Czerwi{\'n}ska, and Maciej Wiznerowicz.
\newblock The cancer genome atlas (tcga): an immeasurable source of knowledge.
\newblock {\em Contemporary oncology}, 19(1A):A68, 2015.

\bibitem{lambert2018human}
Samuel~A Lambert, Arttu Jolma, Laura~F Campitelli, Pratyush~K Das, Yimeng Yin,
  Mihai Albu, Xiaoting Chen, Jussi Taipale, Timothy~R Hughes, and Matthew~T
  Weirauch.
\newblock The human transcription factors.
\newblock {\em Cell}, 172(4):650--665, 2018.

\bibitem{grossman2016toward}
Robert~L Grossman, Allison~P Heath, Vincent Ferretti, Harold~E Varmus,
  Douglas~R Lowy, Warren~A Kibbe, and Louis~M Staudt.
\newblock Toward a shared vision for cancer genomic data.
\newblock {\em New England Journal of Medicine}, 375(12):1109--1112, 2016.

\bibitem{liftover}
Tcga methylation array harmonization workflow.
\newblock Accessed: 2022-05-25.

\bibitem{zhou2018sesame}
Wanding Zhou, Timothy~J Triche~Jr, Peter~W Laird, and Hui Shen.
\newblock Sesame: reducing artifactual detection of dna methylation by infinium
  beadchips in genomic deletions.
\newblock {\em Nucleic acids research}, 46(20):e123--e123, 2018.

\bibitem{liu2009nonparanormal}
Han Liu, John Lafferty, and Larry Wasserman.
\newblock The nonparanormal: Semiparametric estimation of high dimensional
  undirected graphs.
\newblock {\em Journal of Machine Learning Research}, 10(Oct):2295--2328, 2009.

\bibitem{tcga_rnaseq}
Tcga mrna analysis pipeline.
\newblock Accessed: 2022-05-25.

\bibitem{dobin2013star}
Alexander Dobin, Carrie~A Davis, Felix Schlesinger, Jorg Drenkow, Chris
  Zaleski, Sonali Jha, Philippe Batut, Mark Chaisson, and Thomas~R Gingeras.
\newblock Star: ultrafast universal rna-seq aligner.
\newblock {\em Bioinformatics}, 29(1):15--21, 2013.

\bibitem{moore2013dna}
Lisa~D Moore, Thuc Le, and Guoping Fan.
\newblock Dna methylation and its basic function.
\newblock {\em Neuropsychopharmacology}, 38(1):23--38, 2013.

\bibitem{KULIS201027}
Marta Kulis and Manel Esteller.
\newblock 2 - dna methylation and cancer.
\newblock In Zdenko Herceg and Toshikazu Ushijima, editors, {\em Epigenetics
  and Cancer, Part A}, volume~70 of {\em Advances in Genetics}, pages 27--56.
  Academic Press, 2010.

\bibitem{hirasawa2008krab}
Ryutaro Hirasawa and Robert Feil.
\newblock A krab domain zinc finger protein in imprinting and disease.
\newblock {\em Developmental cell}, 15(4):487--488, 2008.

\bibitem{shi2019zfp57}
Hui Shi, Ruslan Strogantsev, Nozomi Takahashi, Anastasiya Kazachenka, Matthew~C
  Lorincz, Myriam Hemberger, and Anne~C Ferguson-Smith.
\newblock Zfp57 regulation of transposable elements and gene expression within
  and beyond imprinted domains.
\newblock {\em Epigenetics \& chromatin}, 12(1):1--13, 2019.

\bibitem{mackay2008hypomethylation}
Deborah~JG Mackay, Jonathan~LA Callaway, Sophie~M Marks, Helen~E White, Carlo~L
  Acerini, Susanne~E Boonen, Pinar Dayanikli, Helen~V Firth, Judith~A Goodship,
  Andreas~P Haemers, et~al.
\newblock Hypomethylation of multiple imprinted loci in individuals with
  transient neonatal diabetes is associated with mutations in zfp57.
\newblock {\em Nature genetics}, 40(8):949--951, 2008.

\bibitem{takahashi2019znf445}
Nozomi Takahashi, Andrea Coluccio, Christian~W Thorball, Evarist Planet, Hui
  Shi, Sandra Offner, Priscilla Turelli, Michael Imbeault, Anne~C
  Ferguson-Smith, and Didier Trono.
\newblock Znf445 is a primary regulator of genomic imprinting.
\newblock {\em Genes \& development}, 33(1-2):49--54, 2019.

\bibitem{chen2019zfp57}
Lie Chen, Xiaowei Wu, Hui Xie, Na~Yao, Yiqin Xia, Ge~Ma, Mengjia Qian, Han Ge,
  Yangyang Cui, Yue Huang, et~al.
\newblock Zfp57 suppress proliferation of breast cancer cells through
  down-regulation of mest-mediated wnt/$\beta$-catenin signalling pathway.
\newblock {\em Cell death \& disease}, 10(3):1--15, 2019.

\bibitem{cheng2022disruption}
Zhaobo Cheng, Renjie Yu, Li~Li, Junhao Mu, Yijia Gong, Fan Wu, Yujia Liu,
  Xiangyi Zhou, Xiaohua Zeng, Yongzhong Wu, et~al.
\newblock Disruption of znf334 promotes triple-negative breast carcinoma
  malignancy through the sfrp1/wnt/$\beta$-catenin signaling axis.
\newblock {\em Cellular and Molecular Life Sciences}, 79(5):1--17, 2022.

\bibitem{sun2022dna}
Dapeng Sun, Xiaojie Gan, Lei Liu, Yuan Yang, Dongyang Ding, Wen Li, Junyao
  Jiang, Wenbin Ding, Linghao Zhao, Guojun Hou, et~al.
\newblock Dna hypermethylation modification promotes the development of
  hepatocellular carcinoma by depressing the tumor suppressor gene znf334.
\newblock {\em Cell death \& disease}, 13(5):1--12, 2022.

\bibitem{ye2018downregulation}
Hui Ye and Meiling Duan.
\newblock Downregulation of foxo6 in breast cancer promotes
  epithelial--mesenchymal transition and facilitates migration and
  proliferation of cancer cells.
\newblock {\em Cancer management and research}, 10:5145, 2018.

\bibitem{yu2020knockdown}
Xixiang Yu, Xixi Gao, Xiaoping Mao, Zhenjing Shi, Bangxuan Zhu, Linqin Xie,
  Shaodan Di, and Limin Jin.
\newblock Knockdown of foxo6 inhibits glycolysis and reduces cell resistance to
  paclitaxel in hcc cells via pi3k/akt signaling pathway.
\newblock {\em OncoTargets and therapy}, 13:1545, 2020.

\bibitem{da2017transcription}
WA~Da~Silveira, PVB Palma, RD~Sicchieri, Rolando~AR Villacis, LRM Mandarano,
  TMG Oliveira, HMR Antonio, Jurandyr Moreira~de Andrade, Valdair~Francisco
  Muglia, SR~Rogatto, et~al.
\newblock Transcription factor networks derived from breast cancer stem cells
  control the immune response in the basal subtype.
\newblock {\em Scientific reports}, 7(1):1--13, 2017.

\bibitem{zou2022ikzf3}
Yan Zou, Bo~Liu, Long Li, Qinan Yin, Jiaxing Tang, Zhengyu Jing, Xingxu Huang,
  Xuekai Zhu, and Tian Chi.
\newblock Ikzf3 deficiency potentiates chimeric antigen receptor t cells
  targeting solid tumors.
\newblock {\em Cancer Letters}, 524:121--130, 2022.

\bibitem{sato2006orphan}
Noriko Sato, Mitsumasa Kondo, and Ken-ichi Arai.
\newblock The orphan nuclear receptor gcnf recruits dna methyltransferase for
  oct-3/4 silencing.
\newblock {\em Biochemical and biophysical research communications},
  344(3):845--851, 2006.

\bibitem{okano1999dna}
Masaki Okano, Daphne~W Bell, Daniel~A Haber, and En~Li.
\newblock Dna methyltransferases dnmt3a and dnmt3b are essential for de novo
  methylation and mammalian development.
\newblock {\em Cell}, 99(3):247--257, 1999.

\bibitem{willis2015enriched}
Scooter Willis, Pradip De, Nandini Dey, Bradley Long, Brandon Young, Joseph~A
  Sparano, Victoria Wang, Nancy~E Davidson, and Brian~R Leyland-Jones.
\newblock Enriched transcription factor signatures in triple negative breast
  cancer indicates possible targeted therapies with existing drugs.
\newblock {\em Meta gene}, 4:129--141, 2015.

\bibitem{madsen2018reparameterization}
Michael~J Madsen, Stacey Knight, Carol Sweeney, Rachel Factor, Mohamed Salama,
  Inge~J Stijleman, Venkatesh Rajamanickam, Bryan~E Welm, Sasi Arunachalam,
  Brandt Jones, et~al.
\newblock Reparameterization of pam50 expression identifies novel breast tumor
  dimensions and leads to discovery of a genome-wide significant breast cancer
  locus at 12q15.
\newblock {\em Cancer Epidemiology, Biomarkers \& Prevention}, 27(6):644--652,
  2018.

\bibitem{pons2005computing}
Pascal Pons and Matthieu Latapy.
\newblock Computing communities in large networks using random walks.
\newblock In {\em International symposium on computer and information
  sciences}, pages 284--293. Springer, 2005.

\bibitem{csardi2006igraph}
Gabor Csardi, Tamas Nepusz, et~al.
\newblock The igraph software package for complex network research.
\newblock {\em InterJournal, complex systems}, 1695(5):1--9, 2006.

\bibitem{gillespie2022reactome}
Marc Gillespie, Bijay Jassal, Ralf Stephan, Marija Milacic, Karen Rothfels,
  Andrea Senff-Ribeiro, Johannes Griss, Cristoffer Sevilla, Lisa Matthews,
  Chuqiao Gong, et~al.
\newblock The reactome pathway knowledgebase 2022.
\newblock {\em Nucleic acids research}, 50(D1):D687--D692, 2022.

\bibitem{korotkevich2021fast}
Gennady Korotkevich, Vladimir Sukhov, Nikolay Budin, Boris Shpak, Maxim~N
  Artyomov, and Alexey Sergushichev.
\newblock Fast gene set enrichment analysis.
\newblock {\em BioRxiv}, page 060012, 2021.

\bibitem{woodfield2007tfap2c}
George~W Woodfield, Annamarie~D Horan, Yizhen Chen, and Ronald~J Weigel.
\newblock Tfap2c controls hormone response in breast cancer cells through
  multiple pathways of estrogen signaling.
\newblock {\em Cancer research}, 67(18):8439--8443, 2007.

\bibitem{eroles2012molecular}
Pilar Eroles, Ana Bosch, J~Alejandro P{\'e}rez-Fidalgo, and Ana Lluch.
\newblock Molecular biology in breast cancer: intrinsic subtypes and signaling
  pathways.
\newblock {\em Cancer treatment reviews}, 38(6):698--707, 2012.

\bibitem{colaprico2016tcgabiolinks}
Antonio Colaprico, Tiago~C Silva, Catharina Olsen, Luciano Garofano, Claudia
  Cava, Davide Garolini, Thais~S Sabedot, Tathiane~M Malta, Stefano~M Pagnotta,
  Isabella Castiglioni, et~al.
\newblock Tcgabiolinks: an r/bioconductor package for integrative analysis of
  tcga data.
\newblock {\em Nucleic acids research}, 44(8):e71--e71, 2016.

\bibitem{lee2001cross}
Adrian~V Lee, Xiaojiang Cui, and Steffi Oesterreich.
\newblock Cross-talk among estrogen receptor, epidermal growth factor, and
  insulin-like growth factor signaling in breast cancer.
\newblock {\em Clinical Cancer Research}, 7(12):4429s--4435s, 2001.

\bibitem{wang2021role}
Bin Wang, Hanfei Guo, Hongquan Yu, Yong Chen, Haiyang Xu, and Gang Zhao.
\newblock The role of the transcription factor egr1 in cancer.
\newblock {\em Frontiers in Oncology}, 11:775, 2021.

\bibitem{lee2015learning}
Jason~D Lee and Trevor~J Hastie.
\newblock Learning the structure of mixed graphical models.
\newblock {\em Journal of Computational and Graphical Statistics},
  24(1):230--253, 2015.

\bibitem{altenbuchinger2019multi}
Michael Altenbuchinger, Helena~U Zacharias, Stefan Solbrig, Andreas
  Sch{\"a}fer, Mustafa B{\"u}y{\"u}k{\"o}zkan, Ulla~T Schulthei{\ss}, Fruzsina
  Kotsis, Anna K{\"o}ttgen, Rainer Spang, Peter~J Oefner, et~al.
\newblock A multi-source data integration approach reveals novel associations
  between metabolites and renal outcomes in the german chronic kidney disease
  study.
\newblock {\em Scientific reports}, 9(1):1--13, 2019.

\bibitem{ambroise2009inferring}
Christophe Ambroise, Julien Chiquet, and Catherine Matias.
\newblock Inferring sparse gaussian graphical models with latent structure.
\newblock {\em Electronic Journal of Statistics}, 3:205--238, 2009.

\bibitem{ma2016network}
Jing Ma, Ali Shojaie, and George Michailidis.
\newblock Network-based pathway enrichment analysis with incomplete network
  information.
\newblock {\em Bioinformatics}, 32(20):3165--3174, 2016.

\bibitem{siahpirani2017prior}
Alireza~F Siahpirani and Sushmita Roy.
\newblock A prior-based integrative framework for functional transcriptional
  regulatory network inference.
\newblock {\em Nucleic acids research}, 45(4):e21--e21, 2017.

\bibitem{zhuang2022augmented}
Yonghua Zhuang, Fuyong Xing, Debashis Ghosh, Farnoush Banaei-Kashani, Russell~P
  Bowler, and Katerina Kechris.
\newblock An augmented high-dimensional graphical lasso method to incorporate
  prior biological knowledge for global network learning.
\newblock {\em Frontiers in genetics}, page 2405, 2022.

\bibitem{wolpert1997no}
David~H Wolpert and William~G Macready.
\newblock No free lunch theorems for optimization.
\newblock {\em IEEE transactions on evolutionary computation}, 1(1):67--82,
  1997.

\bibitem{hastie2009elements}
Trevor Hastie, Robert Tibshirani, Jerome~H Friedman, and Jerome~H Friedman.
\newblock {\em The elements of statistical learning: data mining, inference,
  and prediction}, volume~2.
\newblock Springer, 2009.

\bibitem{chan2017gene}
Thalia~E Chan, Michael~PH Stumpf, and Ann~C Babtie.
\newblock Gene regulatory network inference from single-cell data using
  multivariate information measures.
\newblock {\em Cell systems}, 5(3):251--267, 2017.

\bibitem{marbach2012wisdom}
Daniel Marbach, James~C Costello, Robert K{\"u}ffner, Nicole~M Vega, Robert~J
  Prill, Diogo~M Camacho, Kyle~R Allison, Manolis Kellis, James~J Collins, and
  Gustavo Stolovitzky.
\newblock Wisdom of crowds for robust gene network inference.
\newblock {\em Nature methods}, 9(8):796--804, 2012.

\bibitem{maathuis2010predicting}
Marloes~H Maathuis, Diego Colombo, Markus Kalisch, and Peter B{\"u}hlmann.
\newblock Predicting causal effects in large-scale systems from observational
  data.
\newblock {\em Nature methods}, 7(4):247--248, 2010.

\end{thebibliography}

\includepdf[pages=-]{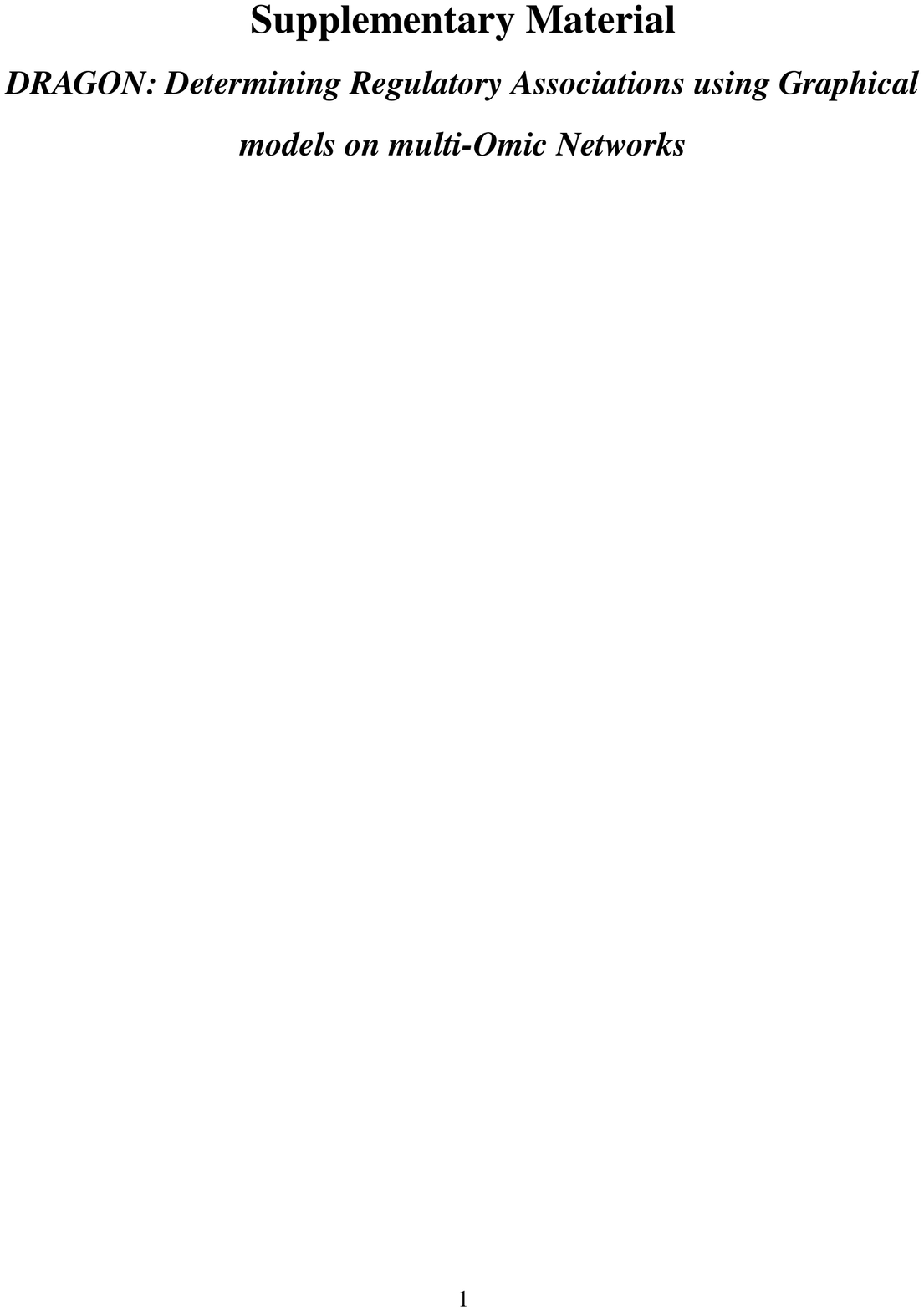} 

\end{document}